\documentclass[usenatbib]{mnras}
\usepackage[T1]{fontenc}
\usepackage{ae,aecompl}
\usepackage{amsmath}
\usepackage{rotating}
\usepackage{graphics}
\usepackage{color}

\newcommand{\ltsimeq}{\la}
\newcommand{\gtsimeq}{\ga}

\newcommand{\msun}{M$_{\odot}$}
\newcommand{\zsun}{Z$_{\odot}$}

\newcommand{\HI}{H{\sc i}}

\newcommand{\ha}{H$\alpha$}

\title[STARBIRDS: Galactic Outflows, SFHs, and Timescales]{Galactic Outflows, Star Formation Histories, and Timescales in Starburst Dwarf Galaxies from STARBIRDS} 
\author[K.~B.~W. McQuinn]{
Kristen.~B.~W. McQuinn,$^{1,2}$\thanks{E-mail: kmcquinn@astro.as.utexas.edu}
Evan D. Skillman,$^{2}$
Taryn N. Heilman,$^{2}$
\newauthor
Noah P. Mitchell,$^{2,3}$
Tyler Kelley$^{2,4}$
\\
$^{1}$The University of Texas at Austin, Department of Astronomy, 2515 Speedway, Stop C1400, Austin, Texas 78712-1205\\
$^{2}$Minnesota Institute for Astrophysics, School of Physics and Astronomy, 116 Church Street, S.E., University of Minnesota, \\Minneapolis, MN 55455\\
$^{3}$Department of Physics and the James Franck Institute, The University of Chicago, 929 East 57th Street, Chicago, Illinois 60637\\ 
$^{4}$Department of Physics and Astronomy, University of California, Irvine, 4129H Frederick Reines Hall, Irvine CA  92697-4575
}

\pubyear{2017}

\begin{document}
\label{firstpage}
\pagerange{\pageref{firstpage}--\pageref{lastpage}}
\maketitle

\begin{abstract}
Winds are predicted to be ubiquitous in low-mass, actively star-forming galaxies. Observationally, winds have been detected in relatively few local dwarf galaxies, with even fewer constraints placed on their timescales. Here, we compare galactic outflows traced by diffuse, soft X-ray emission from {\it Chandra Space Telescope} archival observations to the star formation histories derived from {\it Hubble Space Telescope} imaging of the resolved stellar populations in six starburst dwarfs. We constrain the longevity of a wind to have an upper limit of 25 Myr based on galaxies whose starburst activity has already declined, although a larger sample is needed to confirm this result. We find an average 16\% efficiency for converting the mechanical energy of stellar feedback to thermal, soft X-ray emission on the 25 Myr timescale, somewhat higher than simulations predict. The outflows have likely been sustained for timescales comparable to the duration of the starbursts (i.e., 100's Myr), after taking into account the time for the development and cessation of the wind. The wind timescales imply that material is driven to larger distances in the circumgalactic medium than estimated by assuming short, 5$-$10 Myr starburst durations, and that less material is recycled back to the host galaxy on short timescales. In the detected outflows, the expelled hot gas shows various morphologies which are not consistent with a simple biconical outflow structure. The sample and analysis are part of a larger program, the STARBurst IRregular Dwarf Survey (STARBIRDS), aimed at understanding the lifecycle and impact of starburst activity in low-mass systems. 
\end{abstract} 

\begin{keywords}
galaxies:\ dwarf -- galaxies:\ evolution -- X-rays:\ ISM -- ISM:\ jets and outflows
\end{keywords}

\section{Introduction\label{intro}}
Fundamentally, galaxy evolution is the process of turning gas into stars and subsequently returning energy and enriched material to the galaxy. Feedback from the star formation process can drive outflows of metal-enriched gas which has profound implications for many of the measured properties of galaxies. For example, outflows are invoked to explain the mass-metallicity relation \citep[e.g., MZ relation;][]{Tremonti2004, Lee2006, Berg2012}, the measured metal retention fractions in galaxies \citep[e.g.,][]{Zahid2012, Peeples2014, McQuinn2015f}, and differences in the star formation efficiency of galaxies \citep[e.g.,][]{Kim2017}. 

Dwarfs are a critical testing ground for theories of stellar feedback and provide an opportunity to understand winds and wind models over a larger baseline of galaxy properties. At the low-mass end of the galaxy mass function, shallow potential wells of dwarfs can make them more susceptible to feedback-driven outflows. Analysis of the chemical abundances in dwarfs and theoretical calculations indicate that hot galactic winds in dwarfs are a dominant driver of their supressed metal abundances \citep[e.g.,][]{Garnett2002, Dalcanton2007, Finlator2008, Spitoni2010}. In the latest generation of hydrodynamical simulations \citep[e.g.,][]{Hopkins2012, Sawala2016}, the inclusion of strong stellar feedback and outflows provides a possible solution to discrepancies between predictions from a $\Lambda$CDM cosmology and observations on small mass scales such as the `missing satellite' problem \citep[e.g.,][]{Klypin1999}, the `too big to fail' problem \citep{Boylan-Kolchin2011}, and the cusp-core challenge \citep[e.g.,][]{Navarro1996, Navarro1997, Simon2005, deBlok2010, Walker2011, Oman2015}. 
 
%
\begin{table*}
\begin{center}
\begin{tabular}{ l c c  c  c  c  c l  r c  }
\hline
\hline
Galaxy	& RA 		& Dec			& M$_B$	& Dist.	& A$_V$	& 12$+$log(O/H)	&Principal	& Obs.	& Exp. Time \\
         	& (J2000)		& (J2000) 			&  (mag)  	& (Mpc) 	& (mag) 	&				&Investigator & No.	& (ksec) \\ [0.5ex]
\hline
DDO~165 	& 13:06:24.85 	& $+$67:42:25.0 & 12.92 & 4.81 	& 0.07 	& 7.80			& Jenkins 	&   9537 	&  13.7  \\
                     	&                   	&                      	 &         	&    		&    		&				&    		&10868 	&  10.7 \\ 
NGC~625 &  01:35:04.63 		& $-$41:26:10.3 & 11.65 & 4.21 	& 0.05 	& 8.10			& Skillman &  4746  &  61.1  \\
NGC~1569& 04:30:49.06 		& $+$64:50:52.60 & 11.72 & 3.36	& 1.9 	& 8.19			& Martin   &   782  &  98.0 \\
	                &                     	&                               &    	&    		&		&				& Zezas    &   4745 &  10.2 \\
NGC~4214 & 12:15:39.17 	& $+$36:19:36.8 & 10.21 & 2.95 	& 0.06 	& 8.38			& Zezas    &  4743  &  27.6  \\
                        &                       	&                          &        	&  		&  		&				&    		& 5197  &  29.0 \\
	               &                        	&                          &   	&  		&    		&				& Heckman  &  2030  &  26.8 \\
NGC~4449  & 12:28:11.10	& $+$44:05:37.07 & 9.50 & 4.31 	& 0.05 	& 8.32			& Long  	&  10125 &  15.1  \\
                          &                     	&                           &   	&   		&  		&           			&		& 10875 &  60.2 \\
	                &                       	&                           &   	&   		&   		&				& Heckman  &   2031 &  26.9 \\
NGC~5253  & 13:39:55.96 	& $-$31:38:24.38 & 10.80 & 3.56	& 0.15	& 8.10			& Zezas	& 7153	& 69.1  \\
                         &                    	&                                 &   	&   		&   		&				&          	&  7154  &  67.5 \\
	                &                     	&                                 &   	&   		&   		&				& Heckman  &  2032  &  57.4 \\
\hline	
\end{tabular}
\end{center}
\caption{Summary of galaxy properties and Chandra archival observations for our sample.  Distances are based on the tip of the red giant branch method from \citet{Grocholski2008} for NGC~1569, \citet{Mould2008} for NGC~5253, and \citet{Jacobs2009} for the rest of the sample. A$_V$ are based on Galactic column densities from the \citet{Schlafly2011} recalibration of the \citet{Schlegel1998} dust map.}
\label{tab:galaxies_observations}
\end{table*}

Our empirical perspective on winds in nearby dwarf galaxies is based on the detections of outflows in a surprisingly small sample. Theoretically, winds are predicted to be very hot (T$>10^6$ K), and X-ray observations are necessary to constrain their basic properties such as mass, flux, and energetics. However, due to their transiency and the intrinsically low surface brightness nature of rarefied gas, X-ray detections of winds in dwarf galaxies have been limited to a small sample of galaxies with rapid star formation, known as ``starburst'' galaxies \citep[e.g., NGC~1569, NGC~3077, NGC~4449, NGC~5253, NGC~4214, He2-10, and the extreme system M82;][]{Martin1998, Martin2002, Strickland2000a, Summers2003, Summers2004, Hartwell2004, Ott2005b}. The detections of outflows in these well-cited examples have helped frame the discussion of stellar feedback in dwarfs for more than a decade and provide the primary direct evidence that outflows occur regularly in dwarfs. 

Yet, the physical interpretation of the detected outflows in these landmark papers were based on assumptions about the star formation timescales in the galaxies. Universally, these studies assume that starbursts driving the outflows have ages of 5 Myr based on the presence of young clusters and OB associations, Wolf-Rayet stars, and/or short timescale star formation rate (SFR) indicators such as \ha. However, through the study of the resolved stellar populations in HST imaging, the starburst activity that powers the winds has been measured to persist for 100's Myr $-$ comparable to the dynamical timescales of galaxies \citep{McQuinn2010a, McQuinn2010b}. This suggests that the impact of winds is larger than previously understood and that the winds may also persist for longer than assumed. Furthermore, from the increased attention to stellar-feedback and galactic winds in low-mass galaxies in hydrodynamic simulations \citep[e.g.,][]{Hopkins2012, Muratov2015, Ma2016, vandeVoort2016, Angles-Alcazar2016}, a more detailed, predictive framework of outflows in dwarf galaxies is emerging that can be tested by combining quantified star formation histories (SFHs) and observed outflow characteristics.

Here, we re-examine the existing {\it Chandra} X-ray observations of six dwarf galaxies that also have existing HST imaging of resolved stars and derived SFHs from the STARBurst IRregular Dwarf Survey \citep[STARBIRDS][]{McQuinn2015b}. We also combine observations from multiple {\it Chandra} observing campaigns for greater sensitivity in detecting diffuse X-ray emission. We focus on connecting galactic outflow properties with the measured spatial and temporal properties of the star formation activity and \HI\ distribution. We do not attempt to replicate the detailed analysis in the previous papers on each galaxy, but instead focus on analyzing the X-ray imaging in light of the quantified SFHs. 

Our analysis provides the first empirical measurement of a wind timescale, which we find to be of order 25 Myr.  Our measurements of the energy content of these outflows indicate that they have been sustained for a significant fraction of the starburst duration. This work is part of a larger effort to measure the observed frequency of galactic outflows in low-mass galaxies, and quantify the relationship between star formation activity, outflows, and the location of low-mass galaxies on the MZ relation. 
 
The paper is organized as follows: \S\ref{sec:data} presents the {\it Chandra} X-ray observations, ancillary archival \HI, ultraviolet (UV), and broad-band imaging data, and the relevant previous STARBIRDS results. \S\ref{sec:xray} describes the data reduction including the steps used to remove the point sources in the X-ray data and calibrate the diffuse emission. In \S\ref{sec:morphology} we identify galactic wind features in the individual galaxies. In \S\ref{sec:sfh} we compare the diffuse X-ray luminosity to different star formation timescales based on the SFHs. Our conclusions are summarized in \S\ref{sec:conclusions}.

\section{The Galaxy Sample and Observations}\label{sec:data}
Listed in Table~\ref{tab:galaxies_observations}, the sample consists of six galaxies. All galaxies are part of STARBIRDS, a multi-wavelength study of the lifecycle and impact of starbursts in twenty nearby (D$<$6 Mpc) dwarf galaxies. Existing results from STARBIRDS include SFHs derived from HST imaging of resolved stars \citep{McQuinn2010a}. The six galaxies were selected based on the availability of archival {\it Chandra} data obtained with the Advanced CCD Imaging Spectrometer (ACIS) S-array. In addition, for comparison with the diffuse X-ray emission, all galaxies have deep GALEX Space Telescope UV imaging available from the STARBIRDS program \citep{McQuinn2015b}, and archival VLA maps of the \HI\ \citep{, Cannon2011a, Lelli2014b}.
 
One unifying characteristic of the sample is that all galaxies have experienced recent or on-going bursts of star formation lasting 100's Myr based on SFHs derived from color-magnitude diagrams (CMDs). However, it is important to note that the recent SFHs are varied across the sample, and SFRs also change within each individual galaxy over the lifetime of a burst event. Four systems show on-going bursts, one galaxy (NGC~5253) has declining SFRs over the past 100 Myr, and one galaxy (NGC~625) is a post-burst system with star formation activity returning to its historical average level 65 Myr ago \citep{McQuinn2010b}. In Table~\ref{tab:star_formation}, we summarize the star formation characteristics including the time since the peak SFR, which corresponds to the period of greatest stellar-feedback. The spatial distribution of the recent star formation also varies within the sample, ranging from centrally concentrated in NGC~625, NGC~1569, NGC~5253, and NGC~4449 to more spatially distributed in DDO~165 \citep{McQuinn2012a}.

In the following sections, we describe the suite of observations used in our analysis including the archival {\it Chandra} X-ray observations, archival VLA \HI\ maps, optical catalogs of the stellar populations, and UV and broad-band imaging.

\subsection{X-ray Observations\label{Xray_obs}}
The X-ray data consist of imaging from the {\it Chandra} ACIS-S3 CCD detector. ACIS-S3 has high quantum efficiency (0.8 at 1 keV), high spectral resolution (120 eV), high angular resolution of 1\arcsec\ \citep{Garmire2003}, and a wide field of view (8.\arcmin3$\times$8.\arcmin3). The S3 chip is back-illuminated, which offers a better response at lower energies, making it ideally suited for studying diffuse emission. 

The observations are listed in Table~\ref{tab:galaxies_observations} with the original PI and associated observation number. With the exception of NGC~625, each target was observed more than once in different observing campaigns with overlapping, non-identical fields of view. For these galaxies, the data were reduced individually by epoch, and then reprojected on a common spatial grid and combined for analysis. Combining multiple observing epochs increased the total exposure times (i.e., greater than 100 ks in the cases of NGC~1569, NGC~4449, and NGC~5253) and signal to noise ratios (S/N), enabling the detection of low surface brightness extended X-ray emission with a higher confidence level. The footprints of the combined imaging are larger than the areal coverage of the $HST$ and VLA observations. 

%
\begin{table*}
\begin{center}
\begin{tabular}{lcccccc}
\hline
\hline 	
			&			&			& Peak Burst  		& Time	& 			& \\
			& M$_*$		& sSFR		& SFR 			& since	& Burst		& Concentration \\
	  		& $\times10^6$&$\times10^{-10}$&$\times10^{-3}$ 	& Burst Peak& Duration	& of Stars	 \\
Galaxy  		& (\msun)		&(yr$^{-1}$)	&(\msun\ yr$^{-1}$)	& (Myr)	& (Myr)		& (\%)	 \\
\hline
DDO~165         	& 190		& 4			& 80$\pm$5		& 65		& $>1300\pm500$	& 45	\\ 
NGC~625        	& 260		& 2			& 40$\pm$2		& 450 (65) & $450\pm50$		& 90 \\ 
NGC~1569      	& 700		& 2			& 130$\pm$40		& 65		& $>450\pm50$	& 85  \\ 
NGC~4214     	& 280		& 5			& 130$\pm$40		& 450	&$>810\pm190$	& ... \\ 
NGC~4449     	& 2100		& 5			& 970$\pm$70		& 7.5		&$>450\pm50$		& 72	\\ 
NGC~5253     	& 150		& 6			& 970$\pm$70		& 450	&$>450\pm50$		& 87	\\ 
\hline
\end{tabular}
\end{center}
\caption{Star Formation Properties: Measurements of star formation properties for the galaxies in our sample from \citep{McQuinn2010b}. Col. 2. The cumulative stellar mass in the $HST$ field of view based on the total stellar mass formed over the lifetime of the galaxy and assuming a recycling fraction of 30\% \citep{Kennicutt1994}. Col. 3. Average SFR over the past 10 Myr normalized by the cumulative stellar mass. Col. 4. Peak (not average) SFR of the burst. Col. 5 Time since the peak SFR of the burst. For post-starburst galaxy NGC~625, the second value represents the time since the burst ended. Col. 6. Burst duration. Col. 7 Concentration of the recent star formation activity relative to the stellar disc calculated from the spatial distribution of BHeB to RGB stars \citep[i.e., 1 - BHeB extent/RGB extent;][]{McQuinn2012a}. For NGC~4214, the areal coverage of the $HST$ imaging was insufficient to compare the spatial distribution of the BHeB and RGB stars.}
\label{tab:star_formation}
\end{table*}

\subsection{Star Formation Histories (SFHs)} 
The SFHs were reconstructed from the resolved stellar populations using the numerical CMD fitting technique {\tt MATCH} \citep{Dolphin2002a}. Briefly, {\tt MATCH} uses a initial mass function (IMF) and a stellar evolutionary library to create a series of synthetic simple stellar populations (SSPs) of different ages and metallicities. The synthetic SSPs are modeled using the photometry and recovered fractions of the artificial stars as primary inputs. The modeled CMD that best-fits the observed CMD based on a Poisson likelihood statistic provides the most likely SFH of the galaxy. The SFH solutions were based on a Salpeter IMF \citep{Salpeter1955}, an assumed binary fraction of 35$\%$ with a flat secondary distribution, and the Padua stellar library \citep{Marigo2008}. The temporal resolution of the SFHs depends on the look-back time, with the finest resolution achievable at the most recent times (i.e., $<$1 Gyr). For the purposes of comparing with diffuse X-ray emission, we use the SFRs and stellar mass formed over different timescales within the last 400 Myr from the SFHs. For a full description of the data and applied method, we refer the reader to \citet{McQuinn2010a} and references therein. 

\subsection{VLA 21~cm Maps} 
The spatial distribution of the \HI\ provides an important constraint in evaluating the extent of any hot phase outflows detected in soft X-ray emission. In addition, velocity fields from moment 1 maps of the \HI\ can reveal bulk motions as a result of outflows, providing further context for extended diffuse X-ray emission. All galaxies have existing 21~cm observations from the VLA. We use the publicly available data cubes for five galaxies (NGC~625, NGC~1569, NGC~4214, NGC~4449, NGC~5253) from \citet{Lelli2014b}. These data were blanked below 3$\sigma$ and integrated over all velocity channels to produce to moment 0 maps using AIPS software. We use the moment 0 map for DDO~165 from \citet{Cannon2011a}.

\subsection{GALEX UV and SDSS Broad-Band Imaging} 
Both UV and optical images provide additional context for the spatial extent of both the neutral gas in the 21~cm maps and the hot gas detected in the soft X-ray images. We use the wide field of view GALEX  NUV (1750-2250 \AA) images from \citet{McQuinn2015b}. R-band optical images are from the Sloan Digital Sky Survey (SDSS) data release 12 for DDO~165, NGC~4214, and NGC4449 and from the Digital Sky Survey (DSS) 2 for NGC~625, NGC~1569, and NGC~5253 obtained via the Aladin sky atlas database \citep{Boch2014, Bonnarel2000}.

\section{X-ray Data Reduction}\label{sec:xray}
Data reduction of the {\it Chandra} ACIS-S3 data was carried out in four main stages using {\tt CIAO} (version 4.7), \textit{ACIS Extract} \citep[AE,][]{Broos2010}, and sherpa spectral models \citep{Freeman2001} following the methodology detailed in \citet{Tullmann2011} and \citet{Binder2012}. First, the X-ray data for each observation epoch were separated into energy bands, deflared, and background subtracted. Second, point sources were identified, analyzed, and subtracted, enabling the detection of outflows traced by the diffuse X-ray emission without bias from discrete sources. Third, gas temperatures and X-ray fluxes were estimated by fitting the diffuse X-ray emission with spectral models. Finally, flux maps were made for each observation epoch and combined to create final mosaics of the diffuse soft X-ray. We described each reduction stage in detail below.

\subsection{Pre-Processing and Image Cleaning}\label{clean}
The primary data set from {\it Chandra} is an events file which consists of photon counts in 4 dimensions (two spatial dimensions, time, and energy). Data processing used the \textit{chandra\_repro} tool, which updated the original events file with re-calibrations, identified bad pixels, and constructed aspect solution files describing the orientation of the telescope as a function of time for each observation. The data were divided into soft, medium, and hard energy bands for analysis. For consistency across the sample, energy bands for all observations were uniformly chosen to be 0.35-1 keV (soft), 1-2 keV (medium), and 2-7 keV (hard). We chose a wider soft band than previous X-ray studies on these data \citep[e.g.,][]{Martin2002, Summers2003, Hartwell2004, Ott2005a} to increase the signal to noise ratio in the soft band, where diffuse emission from hot phase outflows is expected. 

Flares in the X-ray data are caused by cosmic rays and their afterglows and manifest as time intervals with unusually high count rates. In cases where flares lead to pixel saturation, the area surrounding the saturated pixels can have low counts and varying afterglow effects. Therefore, the data were filtered to remove time intervals of both higher and lower than average count rates. This was done iteratively using the tool \textit{lc\_sigma\_clip} with a $3\sigma$ inclusion criteria. Cosmic ray afterglows may last several dozen data frames, so time bins were selected to be 144s, 45 times the intrinsic bin size of 3.2 s, to exclude the decaying tails of the afterglows. This stringent criterion increases confidence that the data are clean, which was more important for our science goals than small losses of good exposure time caused by the larger time bins. 

Background subtraction was performed using the background event data set from the {\it Chandra} X-ray Center (CXC). We verified that the background event files were appropriate by statistically comparing them to visually selected empty patches of sky for our sample. For the full energy band, the background event files contained an average of 7$\pm 22$\% fewer counts than our empty sky regions. Likewise, for the soft energy band, the background event files contained an average of 3$\pm 28$\% fewer counts than our empty sky regions. For the final background subtraction of the primary image files, we use the CXC event files normalized by exposure time.

\subsection{Point Source Identification and Subtraction}\label{point_sources}
The background subtracted soft X-ray images include both diffuse emission from the hot gas and emission from individual point sources such as AGN and X-ray binary star systems. Thus, in order to accurately map the morphology of the diffuse soft X-ray emission and measure the total X-ray flux originating solely from the hot gas, we identified and removed compact point sources from the X-ray images. 

Point source identification was a multi-step process. First, point sources were located in each image using the {\tt CIAO} tool \textit{wavdetect}. \textit{wavdetect} correlates adjacent pixels using ``Mexican Hat'' wavelet functions with different scale sizes. Correlations are quantified based on a Poisson signifcance threshold, or \textit{sigthresh}, which identifies the likelihood of a pixel being part of a point source. We used a conservative significance threshold value of $10^{-4}$ to ensure the identification of all real sources. Such a low threshold also identifies a number of false sources (i.e., $\sim$100 per megapixel). In addition, as nearly all of the observations have overlapping fields of view, many point source candidates identified in each image were duplicated in the other source list(s) for a galaxy. Thus, these initial source lists from \textit{wavdetect}, once merged by time bin, energy band, and epoch, required additional filtering to remove false detections and duplicates. 

\begin{table}
\begin{center}
\begin{tabular}{l c c}
\hline
\hline
			&  Final No. 	& L$\rm{_{min}}$	\\ 
Galaxy  		& of Sources	& erg s$^{-1}$	\\
\hline
DDO~165  	& 15  		& 3.0$\times10^{36}$\\
NGC~625 	&  25 		& 2.5$\times10^{35}$\\
NGC~1569 	&  50 		& 3.3$\times10^{35}$\\
NGC~4214  	&  40 		& 1.6$\times10^{35}$\\
NGC~4449  	&  42 		& 9.6$\times10^{35}$\\ 
 NGC~5253	&  51 		& 7.0$\times10^{35}$\\ 
\hline	
\end{tabular}
\end{center}
\caption{Final number of bona fide X-ray point sources confirmed using ACIS Extract and luminosity threshold for detection of point sources.}
\label{tab:galaxies_pointsources}
\end{table}

\begin{table*}
\begin{center}
\setlength{\tabcolsep}{0.03in}
\begin{tabular}{  l  c c c  c c  c  c  c  c  c c c c c}
\hline
\hline
Galaxy  	& N$_{H}$ & Model 	& Reduced & kT$_{1}$ 	& norm$_1$ 	& N$_{H,1}$ & kT$_2$ & norm$_2$	& N$_{H,1}$ 	& Flux		& Flux 		& Lxray$_{soft}$ 	& Lxray$_{soft}^0$\\ 
  		& $\times10^{22}$ & 	& $\chi ^2$ & 		& $10^{-4}$ & $10^{22}$ & $10^{-4}$ &	& $10^{22}$  &$10^{-5}$	& $10^{-14}$ 	& $10^{38}$ 	& $10^{38}$\\
  		&  (cm$^{-2}$)	& 	&	& (keV) 		& (cm$^{-5}$) 	& (cm$^{-2}$)	& (keV) 	& (cm$^{-5}$) 	& (cm$^{-2}$)  	&(cts s$^{-1}$ & (ergs s$^{-1}$ & (erg s$^{-1}$)& (erg s$^{-1}$) \\
		&			&	&	&			&			&			&		&			&		& cm$^{-2}$)	&  cm$^{-2}$) 	&	&\\
\hline
NGC~625 & 0.0084 	& apec	& 0.47	& 0.29$^{+0.06}_{-0.12}$ 	& 0.4	& 0.09 	& -- 		& -- 			& --			&1.2$\pm$1.0	& 1.2 $\pm$1.0	& 0.26$\pm$0.22 	& -- \\
		&  		& mekal 	& 0.47	& 0.29$\pm{0.05}$		& 0.3& 0.01	& -- 		& -- 			& -- 			& 1.4$\pm$0.7 &  1.3$\pm$0.6	& 0.27$\pm$0.13 	& 0.4$\pm$0.1 \\
NGC~1569& 0.359  	& apec 	& 0.33	& 0.47$\pm0.16$		& 12	& 0.66	& 0.14$^{+0.02}_{-0.05}$ & 260& 0.64	& 17$\pm$26	& 19$\pm$29	& 2.6$\pm$ 3.9 	& -- \\
		&  		& mekal 	& 0.33	& 0.68$^{+0.01}_{-0.06}$	& 5	& 0.78	& 0. 27$^{+0.02}_{-0.04}$& 19 	& 0.25	& 8.$\pm$4	& 10$\pm$5	& 1.4$\pm$ 0.6 	& 1.6$\pm$1.9\\
NGC~4214& 0.011 	& apec 	& 0.39	& 0.39$^{+0.02}_{-0.04}$ 	& 20 & 0.45 	& -- 		& -- 			& --			& 10.$\pm$1 	& 11 $\pm$ 7	& 1.1$\pm$ 0.7 	& -- \\
		& 		& mekal 	& 0.36	& 0.30$^{+0.04}_{-0.18}$ 	& 1	& 0.13	& -- 		& -- 			& --			& 5.9$\pm$3.5	& 6.6$\pm$3.7 	& 0.69$\pm$0.39 	& 1.2$\pm$0.1\\
NGC~4449 & 0.0099& apec 	& 0.61	& 0.26$^{+0.01}_{-0.02}$	& 24	& 0.30 	& -- 		& -- 			& -- 			& 35$\pm$5	& 39$\pm$4	& 8.6$\pm$1.0 		& -- \\
		&  		& mekal 	& 0.68	& 0.30$^{+0.01}_{-0.11}$	& 8	& 0.03 	& -- 		& -- 			& --			& 40.$\pm$5	& 42$\pm$5	& 9.3$\pm$1.1		& 12$\pm$0.2\\
NGC~5253& 0.029 	& apec 	& 0.57	& 0.39$^{+0.02}_{-0.03}$ 	& 13 	& 0.49 	& -- 		& -- 			& -- 			& 5.6$\pm$1.6	& 6$\pm$2 	& 0.98$\pm$0.29 	& -- \\
		&  		& mekal	& 0.57	& 0.50$^{+0.04}_{-0.14}$ 	& 1.	& 0.01 	& -- 		& -- 			& -- 			& 5.2$\pm$1.1	& $6.1\pm$1.2 	&0.93$\pm$0.12 	& 1.1$\pm$0.1\\
	\hline	
\end{tabular}
\end{center}
\caption{Summary of X-ray spectral fitting model results. Fixed inputs include the Galactic \HI\ column density in column 2 \citep{Dickey1990, Kalberla2005} and metallicities based on the oxygen abundances listed in Table~\ref{tab:galaxies_observations}. Fitted outputs include plasma temperatures, normalizations, and intrinsic absorbing \HI\ column densities (N$_{H,1}$ and N$_{H,2}$), and measured fluxes and luminosities for the diffuse soft (0.35-1.0 keV)  X-ray emission. The final column lists the unabsorbed X-ray luminosities from the mekal spectral model in the 0.35 - 1.0 keV energy range with point sources removed. DDO~165 is not detected in diffuse emission and is therefore omitted from the Table.}
\label{tab:specfits_fluxes}
\end{table*}

We used the tool AE to filter the merged list of possible point sources into a final list of sources to be removed from the images. For each source listed from \textit{wavdetect}, AE measures the flux and assigns a probability that no source actually exists (a probability no source or pns value). The pns value is the Poisson probability that all of the counts within the extraction region are background noise. Initially, sources with a pns above 10$^{-2}$ were rejected, after visual inspection of the image(s) confirmed that there were no obvious sources at the location. The AE analysis and visual inspection was iteratively repeated on modified source lists with more stringent pns values of 10$^{-3}$, 10$^{-4}$, and finally 4x10$^{-6}$ (corresponding to a 4.5$\sigma$ detection). Finally, AE was run on the final source list to determine the spatial extent and background subtracted energy flux of each source. The measured energy fluxes were used to estimate the luminosity detection limits following the method of \citet{Broos2011} and adopting the distances listed in Table~\ref{tab:galaxies_observations}. The final number of point sources per galaxy and luminosity thresholds for detection are listed in Table~\ref{tab:galaxies_pointsources}. 

The final AE output contains irregular polygon regions for each point source that nominally enclose 90$\%$ of a source's flux. The regions are conservatively estimated to avoid flux contamination by nearby sources in crowded regions. To ensure full point source subtraction, we used these regions only as a guide and fit each point source by eye with an ellipse. Based on these ellipses, the {\tt CIAO} \textit{roi} and \textit{split roi} tools were used to define the final regions enclosing each source and background annuli. The regions were masked with pixel values interpolated from the surrounding background annuli to create maps of diffuse emission. 

Finally, the diffuse emission maps were smoothed using the {\tt CIAO} \textit{csmooth} tool, which adaptively smoothes using a fast fourier transform algorithm while preserving the total number of counts in each image. Smoothing was done using a circular Gaussian kernel of varying scale, increased until the total number of counts under the kernel exceeded the expected number of background counts in the kernel area. Once smoothed, the diffuse maps were inspected to ensure that the subtracted point sources were fully removed without introducing morphological changes to the emission. One galaxy, DDO~165, is not detected in the diffuse emission maps.

\subsection{Flux Measurements}\label{flux}
We measure the fluxes in the 0.35 - 1.0 keV energy band from the event files using using Sherpa spectral models. After masking the detected point sources in the event files, we extracted spectra within a region defined by $\sim1.5-2\times$ the optical diameter based on the 25 mag arcsec$^{-2}$ isophote in the B-band (i.e., D$_{25}$). We confirmed that these regions encompassed the area enclosed by $2\sigma$ contours of soft X-ray emission. We also extracted spectra from regions outside the galaxies to provide a measurement of the background. Note that while we have masked all identified point sources, unresolved compact sources may be present and contributing to the measured fluxes. While the majority of compact sources are expected to have a harder spectrum that the lower 0.35 - 1.0 keV energy band of interest here, \citet{Mineo2012} estimate that high mass X-ray binaries (HMXBs) contribute an average of 37\% to fluxes measured in a slightly higher 0.5 - 2.0 keV energy range. The largest fractional contribution of compact sources is in the 1.0 - 2.0 keV range, with a smaller $\sim10$\% contribution to fluxes below 1.0 keV. Thus, we conclude that while our measured fluxes are a formal upper limit for emission from the diffuse component, contributions from unresolved compact sources do not significantly impact our fluxes and are within the uncertainties of fluxes measured from the spectral model fits. 

We fit the background-subtracted extracted spectra with two different spectral models in \textit{sherpa} \citep{Freeman2001}, namely the Astrophysical Plasma Emission Code ({\sc apec}) thermal plasma \citep[{\it xsapec};][]{Smith2001, Raymond1977}, and the {\sc mekal} thermal plasma model \citep{Mewe1985, Mewe1986, Kaastra1993}. Both of these spectral models assume either a one or two temperature plasma and account for foreground extinction and metallicity-dependent internal extinction. To estimate the foreground extinction, we used the \textit{sherpa} model \textit{xsphabs} and measurements of the Galactic gas column densities based on the \citet{Schlafly2011} recalibration of the \citet{Schlegel1998} dust map. To estimate the internal extinction, we used the sherpa model \textit{xsvphabs} and adopted the gas-phase oxygen abundances in the sample listed in Table~\ref{tab:galaxies_observations}. 

Table~\ref{tab:specfits_fluxes} lists the model parameters, absorbing column densities of Galactic hydrogen, and outputs. The model outputs were averaged for each galaxy and used to create weighted instrument maps needed to produce final diffuse emission maps. Note that soft X-ray emission is thought to be primarily produced by shocked material in the interacting regions between SN ejecta and the ambient gas, rather than by the wind fluid itself \citep[e.g.,][]{Suchkov1994, Strickland2000a}. From simulations, this thermal soft X-ray emission is thought to represent only $<10$\% of the total energy in a multi-phase outflow \citep{Strickland2000a}, suggesting a simple one or two temperature spectral model is inadequate to physically describe the X-ray emitting gas. However, given the low photon counts and lack of additional constraints from the multi-phase gas in the outflow, we proceed with this simplistic temperature model with the caveat that the outflows are likely more complex.

Table~\ref{tab:specfits_fluxes} lists the source flux in photons, the energy flux in ergs, and the luminosity values assuming the distances in Table~\ref{tab:galaxies_observations} for each spectral model. For nearly all galaxies, the data were fit with single temperature {\sc apec} and {\sc mekal} models. The exception is NGC~1569 where a single temperature spectral model was poorly fit with higher residuals and higher uncertainties. In this case, we used two temperature spectral models. Also included in Table~\ref{tab:specfits_fluxes} are the unabsorbed soft, diffuse X-ray luminosities which take into account both the Galactic and internal absorption estimated from the spectral model fits. For simplicity, we adopt the unabsorbed luminosities from the {\sc mekal} model in subsequent analysis, which is the most commonly used model in the literature for similar analysis \citep[e.g.,][]{Ott2005b, Hartwell2004}. The X-ray luminosities are compared with the SFHs in \S\ref{sec:sfh} below.

\subsection{Diffuse Emission Maps}\label{maps}
The diffuse, point-source and background subtracted X-ray images in units of counts pixel$^{-1}$ were converted to calibrated flux images in units of erg cm$^{-2}$ s$^{-1}$ pixel$^{-1}$ using an instrument map and an effective exposure map. The instrument map provides the instantaneous effective area across the field of view and accounts for the quantum efficiency (QE) of the detectors and spatial non-uniformities of the detector array, mirror vignetting, and bad pixels. The exposure map is the product of the instrument map and the aspect histogram, which gives the time observed for each part of the sky. 

Because the effective area of the instrument maps is energy dependent, we weighted the instrument maps using the Sherpa spectral models. The final weighted instrument maps give the effective area of the telescope mirror projected onto the detector surface with the detector quantum efficiency (QE), as a function of photon energy. The instrument maps are convolved with the aspect histogram of the observations to create exposure maps. Flux calibrated, extinction corrected images were made by dividing the filtered, background subtracted, diffuse counts images by the exposure map, to yield a surface brightness image with units of photons cm$^{-2}$ s$^{-1}$ pixel$^{-1}$.

Finally, these flux-calibrated images were reprojected onto a custom grid large enough to encompass the combined footprint of the multiple observing campaigns using the \textit{reproject\_image\_grid} tool. The reprojected images were averaged by pixel to create surface brightness maps. The maps provide a measure of the spatial extent and morphology of the soft, diffuse X-ray emission in the galaxies.

\begin{figure*}
\includegraphics[width=0.98\textwidth]{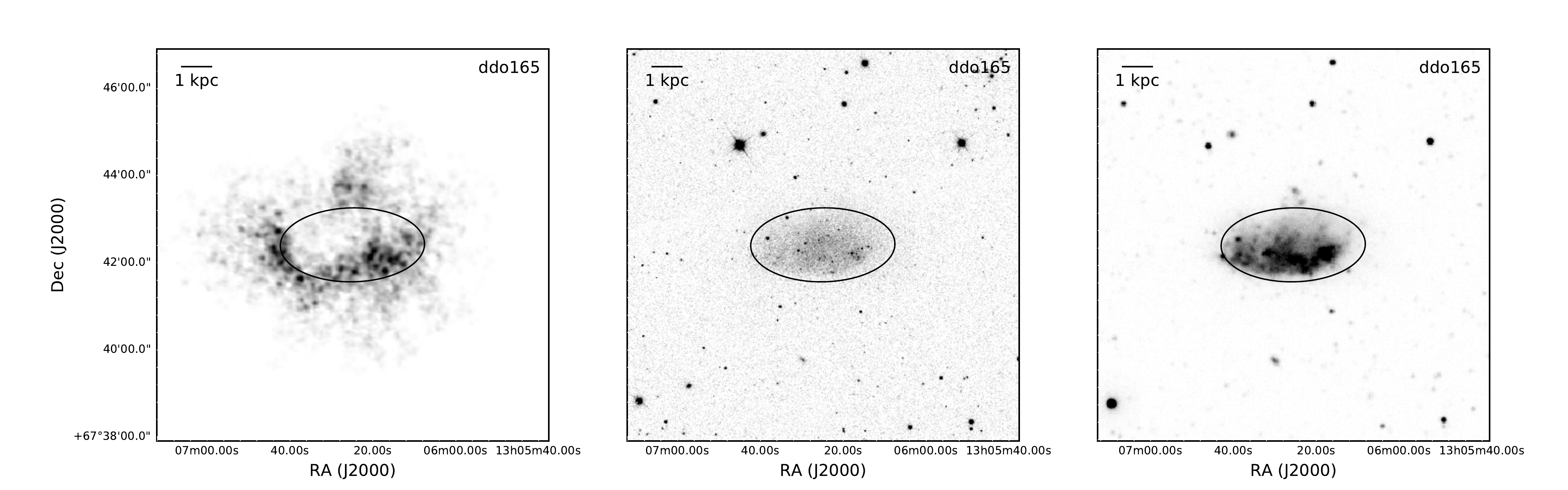}
\caption{DDO~165. {\it Left:} 21 cm neutral hydrogen emission map. {\it Middle:} DSS R-band optical image. {\it Right:} GALEX NUV image. The black ellipse represent the main optical disk of the galaxy measured by the 25 m$_B$ arcsec$^{-2}$ isophote. No diffuse X-ray emission was detected.}
\label{fig:ddo165_xraycontour}
\end{figure*}

\begin{figure*}
\includegraphics[width=0.98\textwidth]{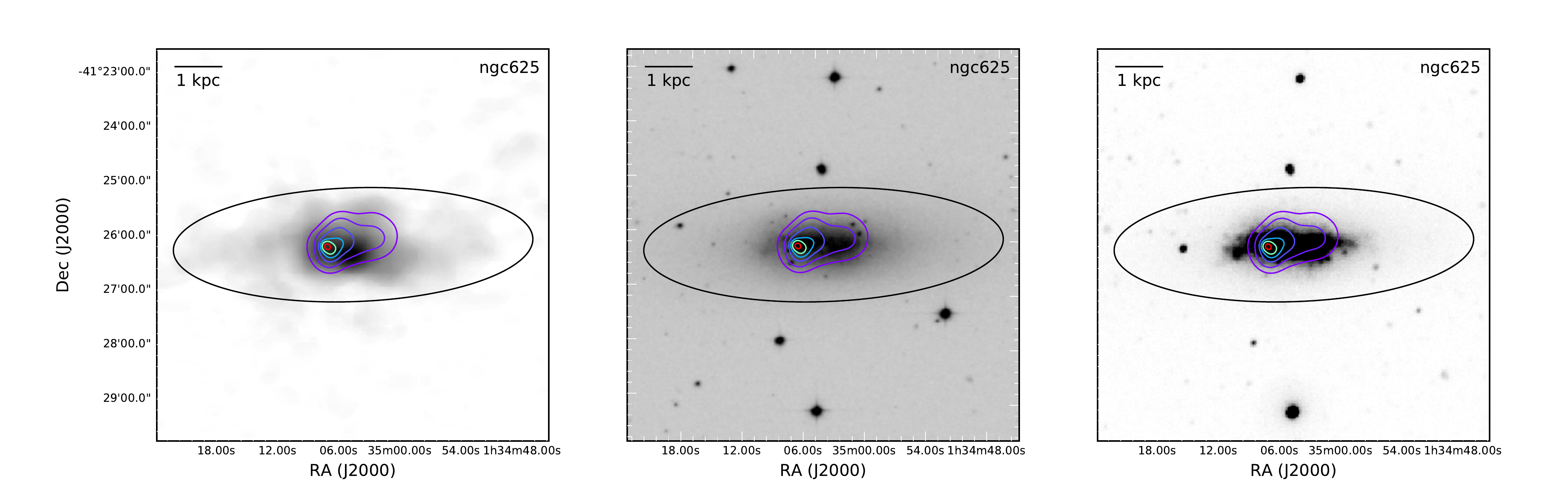}
\caption{NGC~625. {\it Left:} 21 cm neutral hydrogen emission map. {\it Middle:} DSS R-band optical image. {\it Right:} GALEX NUV image. Each image is overlaid with contours of soft X-ray emission corresponding to 2, 4, 8, 16, 32, and 64 $\sigma$. The black ellipse represent the main optical disk of the galaxy measured by the 25 m$_B$ arcsec$^{-2}$ isophote. Extended diffuse X-ray emission is detected around the star forming disc of the galaxy out to the edges of the gaseous disc along the minor axis.}
\label{fig:ngc625_xraycontour}
\end{figure*}

\section{Morphology of the Diffuse Soft X-ray Emission}\label{sec:morphology} 
Figures~\ref{fig:ddo165_xraycontour}$-$\ref{fig:ngc5253_xraycontour} present contours of the diffuse, soft X-ray emission overlaid on \HI, R-band, and NUV images for each galaxy. Black ellipses outline the extent of the main stellar body of the galaxies at the B-band surface brightness level of 25 mag arcsec$^{-2}$ \citep[HyperLeda;][]{Muratov2014}. The morphology of the X-ray emission is varied and chaotic, with little evidence of the biconical structure seen in more energetic outflows such as the M~82 system.

Each galaxy is discussed below comparing the morphology of the diffuse X-ray emission with the neutral gas distribution, stellar disc, and distribution of young, UV-bright populations. We include comparisons from the previous studies using the {\it Chandra} data. As we combine observations from multiple {\it Chandra} programs and use a wider energy band to define the soft X-ray energy band, there are additional features in our diffuse emission maps not detected in previous studies. We make note of these differences below, in particular for NGC~4449 and NGC~5253. Measurements of the angular extent of the soft X-ray emission are provided in Table~\ref{tab:xray_mech}. 
 
\subsection{DDO~165} 
From Figure~\ref{fig:ddo165_xraycontour}, no diffuse X-ray emission is detected within the main disc of the galaxy. DDO~165 is the farthest galaxy in the sample and also has the shortest combined exposure time of 24 ksec, making these X-ray observations the least sensitive of the sample. However, based on deep H$\alpha$ imaging, the warm ionized gas is contained with the \HI\ disc (McQuinn et al.\ in prep) suggesting that it is unlikely that DDO~165 has a hot gas outflow of significance. These observations do not rule out a starburst-driven galactic wind in the recent past; modelling of the energetics of star formation suggest it experienced a large scale blow-out of the \HI\ disc in the last $\ltsimeq100$ Myr \citep{Cannon2011a, Cannon2011b}. 

\subsection{NGC~625} 
From Figure~\ref{fig:ngc625_xraycontour}, the strongest X-ray emission is highly concentrated, but resolved, in the center of the starburst. Diffuse soft X-ray emission surrounds this region and extends above and below, reaching close to the edge of the \HI\ disc in projection in the south. Note that while the data for NGC~625 are archival, the X-ray imaging was not previously published. The \HI\ kinematics are highly disturbed in the center of the galaxy and the complex \HI\ velocity structure at larger radii are consistent with a significant outflow of neutral hydrogen \citep{Cannon2004}. \ha\ imaging of NGC~625 shows low surface brightness extraplanar features above the central starburst in the same area as the detected X-ray emission including a complete loop with a defined edge \citep{Meurer2006}. 

\begin{figure*}
\includegraphics[width=0.98\textwidth]{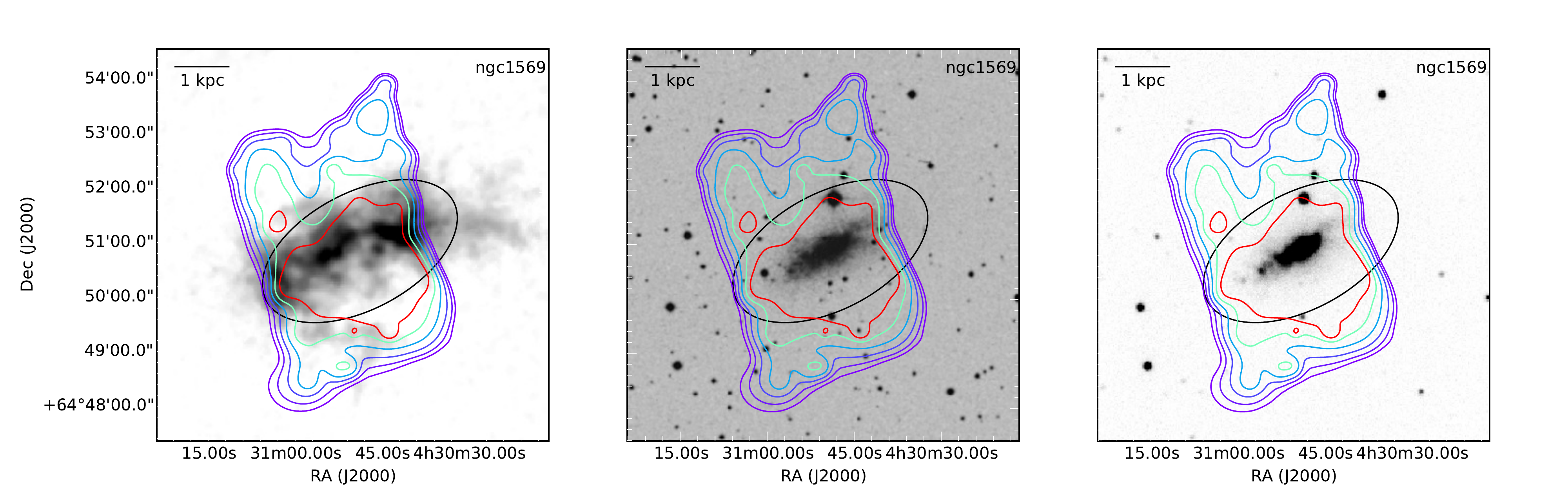}
\caption{NGC~1569. {\it Left:} 21 cm neutral hydrogen emission map. {\it Middle:} DSS R-band optical image. {\it Right:} GALEX NUV image. Each image is overlaid with contours of soft X-ray emission corresponding to 2, 4, 8, 16, 32, and 64 $\sigma$. The black ellipse represent the main optical disk of the galaxy measured by the 25 m$_B$ arcsec$^{-2}$ isophote. Diffuse soft X-ray emission traces a large-scale outflow along the minor axis of the galaxy.}
\label{fig:ngc1569_xraycontour}
\end{figure*}

\begin{figure*}
\includegraphics[width=0.98\textwidth]{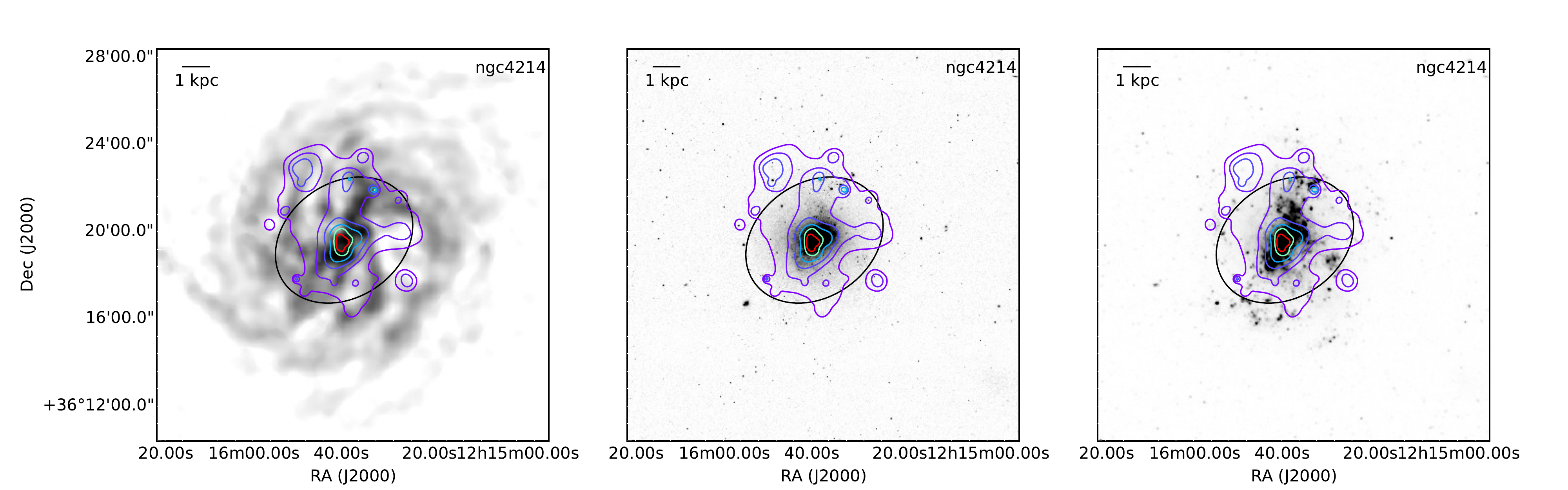}
\caption{NGC~4214. {\it Left:} 21 cm neutral hydrogen emission map. {\it Middle:} DSS R-band optical image. {\it Right:} GALEX NUV image), each overlaid with contours of soft X-ray emission corresponding to 2, 4, 8, 16, 32, and 64 $\sigma$. The black ellipse represent the main optical disk of the galaxy measured by the 25 m$_B$ arcsec$^{-2}$ isophote. The diffuse soft X-ray emission is extended in a number of regions outside the star forming disc and coincides with a low density cavity in the \HI\ to the west.}
\label{fig:ngc4214_xraycontour}
\end{figure*}

\subsection{NGC~1569} 
From Figure~\ref{fig:ngc1569_xraycontour}, there is clear evidence of extended diffuse soft X-ray emission along the minor axis of the galaxy. In contrast to a biconical wind structure, such as observed in M~82 \citep[e.g.,][]{Bland1988}, the outflow is across the full minor axis of NGC~569, with a ``boxy'' morphology. The edges of the outflow suggest they are pressure-bound. The extent of the X-ray emission exceeds both the optical and the \ion{H}{1} disks as far as $\sim4.8$\arcmin, or 3.1 kpc. The strongest X-ray emission is concentrated around the center star forming disc of the galaxy with an extension to the southwest in a region with lower \HI\ column densities. In the sample, NGC~1569 has the strongest evidence of a large-scale hot outflow. This has been noted previously by several studies including the original analysis of the X-ray data \citep{Martin2002}. \ha\ imaging and kinematics of the outflow show faint ionized filaments in the halo with expansion velocities of order $\sim100$ km s$^{-1}$ \citep{Martin1998, Westmoquette2008}. \HI\ kinematics show highly disturbed gas in the center of the galaxy with no detectable rotation. Outside the center region the gas dispersion is  $\sim2 \times$ higher than typical dwarf galaxies \citep{Stil2002}. \HI\ was also detected at larger radii with discrepant, high velocities relative to the bulk of the gas; this is thought to be outflowing gas, although inflowing gas has not been ruled out \citep{Stil2002}.

\begin{figure*}
\includegraphics[width=0.98\textwidth]{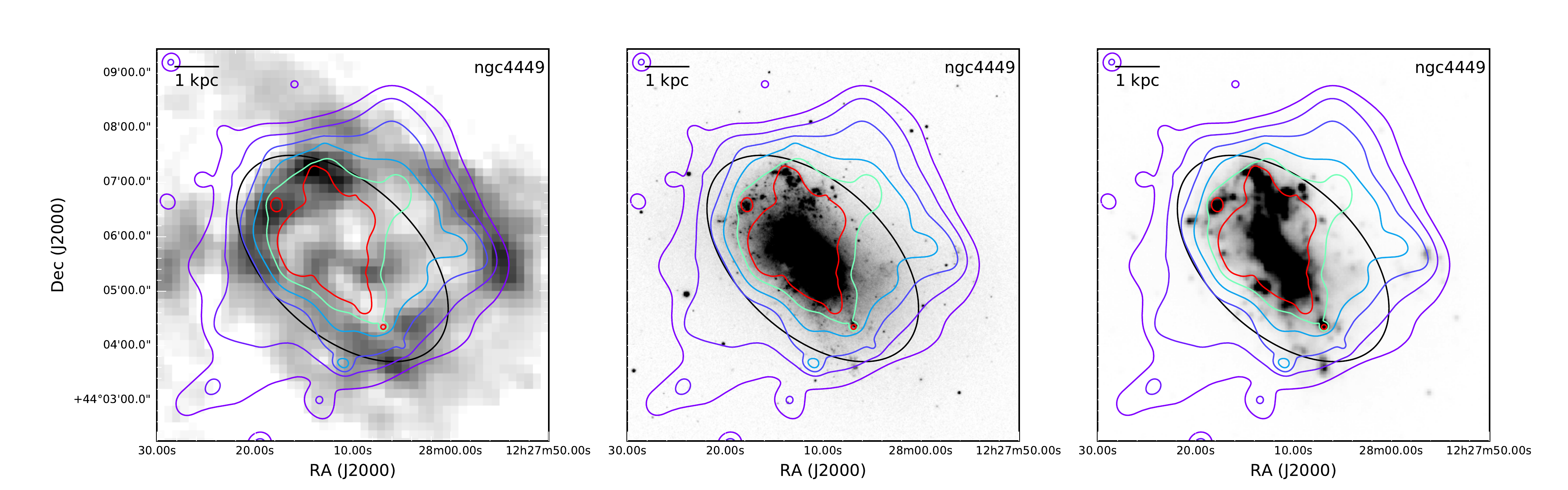}
\caption{NGC~4449. {\it Left:} 21 cm neutral hydrogen emission map. {\it Middle:} DSS R-band optical image. {\it Right:} GALEX NUV image), each overlaid with contours of soft X-ray emission corresponding to 2, 4, 8, 16, 32, and 64 $\sigma$. The black ellipse represent the main optical disk of the galaxy measured by the 25 m$_B$ arcsec$^{-2}$ isophote. The diffuse soft X-ray is widely extended along the minor axis of the galaxy with the strongest evidence of an outflow in the southeast.}
\label{fig:ngc4449_xraycontour}
\end{figure*}

\begin{figure*}
\includegraphics[width=0.98\textwidth]{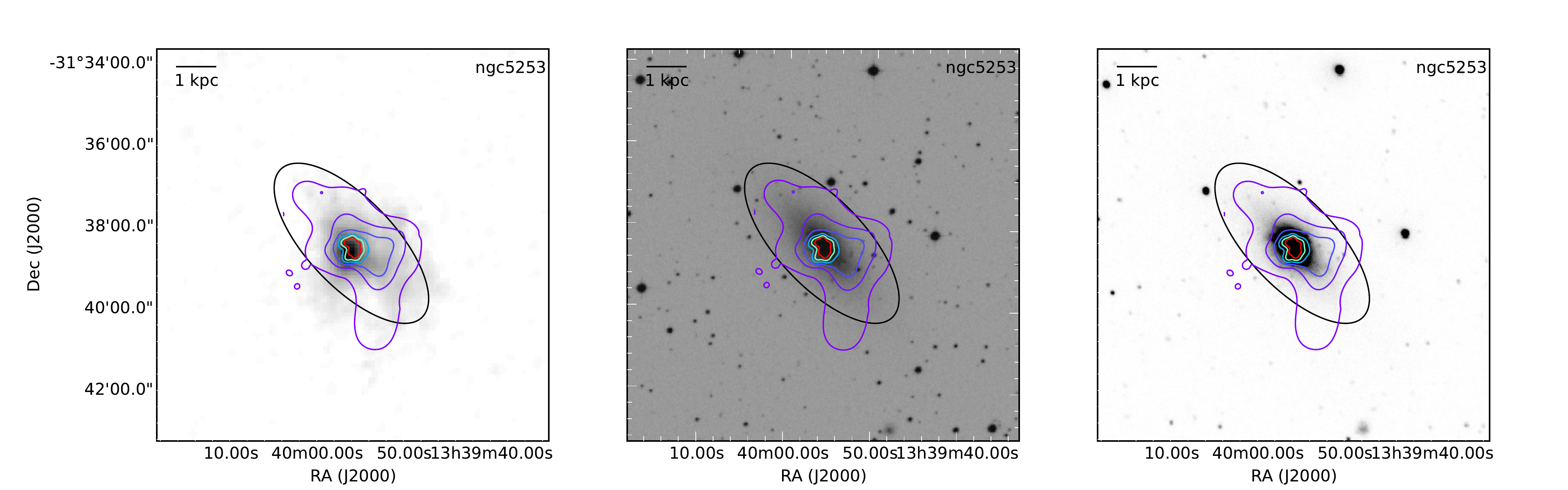}
\caption{NGC~5253. {\it Left:} 21 cm neutral hydrogen emission map. {\it Middle:} DSS R-band optical image. {\it Right:} GALEX NUV image. Each image is overlaid with contours of soft X-ray emission corresponding to 2, 4, 8, 16, 32, and 64 $\sigma$. The black ellipse represent the main optical disk of the galaxy measured by the 25 m$_B$ arcsec$^{-2}$ isophote. Diffuse soft X-ray surrounds the central region and extends past the \HI\ disc in the south and, to a lesser extent, in the west.}
\label{fig:ngc5253_xraycontour}
\end{figure*}

\subsection{NGC~4214}  
Figure~\ref{fig:ngc4214_xraycontour} shows the diffuse soft X-ray emission has a complex morphology. The strongest X-ray emission is in the center of the galaxy, corresponding to a region known to contain expanding bubbles of ionized gas with velocities reaching $\sim100$ km s$^{-1}$ \citep{Martin1998}. Outside this central region, the emission can be characterized by three main components. The first is an elongation in the north-south axis which is aligned with the distribution of young, UV-bright stars seen in the right panel of Figure~\ref{fig:ngc4214_xraycontour}. The second includes extended emission to the west that is much broader and roughly perpendicular to the north-south axis. This irregularly shaped region is coincident with cavities in the \HI\ with lower column densities and are also coincident with low surface brightness ionized gas detected in deep \ha\ imaging (McQuinn et al. in prep).  A previous comparison of the soft X-ray emission with \ha\ imaging in NGC~4214 did not detect ionized gas in this region \citep{Hartwell2004}, but deeper \ha\ observations reveal the presence of warm gas with a similarly irregular morphology. The lack of a sharp boundary in this region suggests that the bubble has ruptured. If directed along the line of sight, this material would constitute an outflow out of the \HI\ disc.  

The third component is the most significant diffuse soft X-ray extension. It lies in the northeast reaching $\sim4.5\arcmin$ (3.4 kpc) from the center of the galaxy. Hot gas is moving out from the main disc of the galaxy in this region, but seen in projection, the emission is still contained within the \HI\ gas. Similarly to the western feature, whether this material will escape from the gaseous disc is difficult to determine given the viewing angle of NGC~4214. The higher intensity X-ray emission (8$\sigma$ contour) is less pronounced in an image smoothed with a fixed-sized Gaussian kernel, and is likely an unresolved background source that was not fully subtracted. Regardless, there is a clear detection of hot gas transporting material outside the main star-forming disc in this region, which has direct implications for the mixing of newly synthesized material to larger radii in the galaxy.

\subsection{NGC~4449}
In Figure~\ref{fig:ngc4449_xraycontour}, the strongest soft X-ray emission is coincident with the central star forming region of the galaxy. Hot gas surrounds this center with the most broadly extended emission along the minor axis reaching 5.5 kpc outside the main \HI\ disc. In the southeast, the diffuse X-ray emission has a finger-like extension into a region of lower \HI\ column density. Note that the \HI\ morphology and kinematics of NGC~4449 are complex with an extended \HI\ arm structure around the central disc, likely influenced by tidal interactions with two nearby companions \citep{Hunter1998, Martinez-Delgado2012, Rich2012}. In the northwest, the X-ray emission also extends past the \HI\ disc, but within the \HI\ arm structure in projection. Comparison with deep \ha\ emission shows that these two X-ray features along the minor axis are also detected in the warm ionized gas phase (McQuinn et al. in prep). 

The morphology of the southeast region is consistent with an outflow from a ruptured superbubble. In previous works, \citet{Summers2003} compared the X-ray morphology to that of the ionized gas traced by \ha\ emission and concluded that the northwest X-ray was the most likely indication of outflowing hot gas. However, the southeast finger in X-ray was not as extended, presumably due to the narrower energy range used for the soft X-ray band and shorter integration times for both the X-ray and \ha\ data. Here, we find that the diffuse soft X-ray emission in the southeast is the clearest indication of an outflow of hot gas in NGC~4449.

\subsection{NGC~5253} 
From Figure~\ref{fig:ngc5253_xraycontour}, the strongest soft X-ray emission is in the center of the galaxy where the starburst is concentrated. There are two other notable outflow features. The first is the extended emission to the west and east, along the minor axis of the galaxy. The western emission is consistent with a bubble of hot gas extending past the edge of the \HI\ disc and corresponds to the previously noted ruptured bubble of warm ionized gas seen in \ha\ \citep{Marlowe1995, Martin1998}. \citet{Summers2004} suggest this gas may be expanding due to higher pressure and ambient interstellar medium (ISM) density of the dust lane bisecting the nuclear region to the east. This is consistent with the identification of an \HI\ feature in the  west that is thought to contain as much as $5\times10^6$ \msun\ in neutral hydrogren \citep{Kobulnicky1995, Kobulnicky2008}. The second feature is in the south and shows hot gas outside the optical disc and in a region of lower \HI\ density reaching $\sim$2.4 kpc from the center. A similar but less extended feature is detected in \ha\ in previous studies \citep{Marlowe1995, Martin1998, Summers2004}. Note there is neutral hydrogen in projection between the eastern and southern X-ray features that is highly kinematically disturbed and thought to be inflowing gas \citep{Kobulnicky2008, Lopez-Sanchez2012}.

\section{Outflows and Timescales: Comparison of X-ray Emission and SFHs}\label{sec:sfh}
The standard picture for the formation of an outflow is hierarchical. In a simplified framework, stellar winds and radiation pressure from young stars carve cavities in the ISM. Energy injected from SNe into the cavities create bubble-like structures that expand outward, and possibly merge with similar structures around neighboring star-forming regions. If powered by enough energy, these merged bubbles or ``superbubbles'' can reach the edge of an \HI\ disc and ``rupture'' to form an outflow. The full process depends significantly on the local star formation activity and ISM conditions including complex properties such as turbulence and magnetic fields. Nonetheless, the overall hierarchical framework is supported by the ubiquitously observed bubble and superbubble structures in actively star forming regions in low-mass galaxies \citep[e.g.,][]{Heckman1995, Martin1998, Martin2002, Summers2003} and the holes seen in the neutral hydrogen of dwarf galaxies \citep[e.g.,][]{Ott2001, Simpson2005, Cannon2011a, Warren2011}.

There are a number of timescales associated with this general scenario including (1) the star formation event, (2) the development of bubbles and superbubbles, (3) the formation of an outflow (i.e., when the bubbles break through the \HI\ disc), and (4) the longevity of an outflow. Here, we present a brief overview of existing constraints on the first three timescales, and use the SFHs to provide additional timescale measurements and a constraint on the longevity of a wind.

\subsection{Formation of an Outflow}
In previous studies of winds in starburst dwarf galaxies, a timescale of 5 Myr was generally assumed for the star formation activity that powers the outflows, based on star formation tracers such as the presence of OB associations, H$\alpha$ emission, and Wolf-Rayet stars \citep[e.g.,][]{Strickland2000a}. The formation of an extended outflow begins with the first generation of SNe in such a star-forming region based on predictions from hydrodynamical simulation \citep{Strickland2000a, Hopkins2012, Onorbe2015, Muratov2015, Kim2017}. 

Observationally, the timescales for bubble development and outflow formation are often determined based on dynamical arguments. The edges of expanding bubble-like structures are limb-brightened and seen as shells in H$\alpha$ imaging. Spectroscopically measured velocities of the H$\alpha$ emission can be compared with the projected physical scales of the bubbles and star forming regions to yield dynamical ages of the structures. Across multiple studies, shell velocities range from 20-400 km/s with projected physical sizes up to $\sim$1.4 kpc which yield dynamical ages of order 10 Myr, and no dynamical ages longer than 20 Myr \citep{Heckman1995, Martin1998, Martin2002, Summers2003}. One interpretation of the maximum $\sim$20 Myr dynamical age is that bubbles larger than this must have already broken through the \HI\ disc and ``ruptured'', creating an outflow from the galaxy \citep{Martin1998}. After an outflow has developed, dynamical arguments break down and cannot be used to estimate the longevity of a feedback-driven wind. 

The typical dynamical age of $\sim10$ Myr for larger shells can be compared with the star formation timescales. As the shells are primarily photoionized by massive stars which have lifetimes $<10$ Myr, this indicates that the star formation events powering the bubble formation must consist of multiple generations of massive stars. This is in conflict with the assumed 5 Myr timescales for the star formation activity typically associated with driving the outflows. Logically, the star formation responsible for creating a wind cannot be younger than the wind itself. Yet, few alternative star formation measurements have been used to measure the activity over a longer period of time. The implied assumption is that the activity traced by the star formation indicators must have been on-going and sustained over longer periods of time.  

The argument for elevated star formation activity occurring over longer timescales is supported by measurements from the SFHs. In contrast to the short $5-10$ Myr timescales, analysis of CMDs show quantitatively that the galaxies have experienced on-going, elevated star formation activity for significantly longer, with burst durations $\gtsimeq$450 Myr (see Table~\ref{tab:star_formation}). In addition, the SFRs are variable and peak $\sim65-450$ Myr ago in five galaxies, indicating that the strongest star formation events powering the outflows occurred on timescales not probed by recent star formation indicators. This highlights that the X-ray observations of winds are tracing only the present-day activity, with little information on the history or age of the outflows. Many of the previous studies note the uncertainties in the age of the outflows based on the unknown past star formation activity \citep[e.g.,][]{Summers2003, Summers2004, Hartwell2004}. 

In a short, 5 Myr starburst paradigm, the winds can be thought of as an extended ``gasp'' of material and energy leaving a galaxy. For longer, more realistic star formation timescales, the duration of the outflows should be comparable to the duration of the starbursts, after taking into account the time for the development and cessation of the wind. This is consistent with expectations from simulations that show outflows and diffuse soft X-ray emission track the star formation activity in low-mass galaxies \citep{vandeVoort2016}. Outflows sustained over time periods of 100's Myr would then provide a continual source of heating and material to the CGM, diminishing or simply slowly the re-accretion rate of gas back to the host galaxy. Gas re-accretion timescales in low-mass galaxies are predicted to be of order 300 Myr \citep{Angles-Alcazar2016}, but it is unclear how this process proceeds when occurring simultaneously with an outflow ongoing over comparable timescales. Assuming the outflows are preferentially enriched with metals, this also implies that a significant amount of the metals produced in starburst events is not available to be recycled in the next generation of star formation, supporting the similar stellar and gas metallicities predicted as a function of time \citep[e.g.,][]{Muratov2017}. However, given that 50-95\% of the ejected material is predicted to be recycled into low-mass galaxies over cosmic time \citep{Angles-Alcazar2016}, it is possible that many of metals may eventually return to the host galaxy.

\begin{figure}
\includegraphics[width=0.48\textwidth]{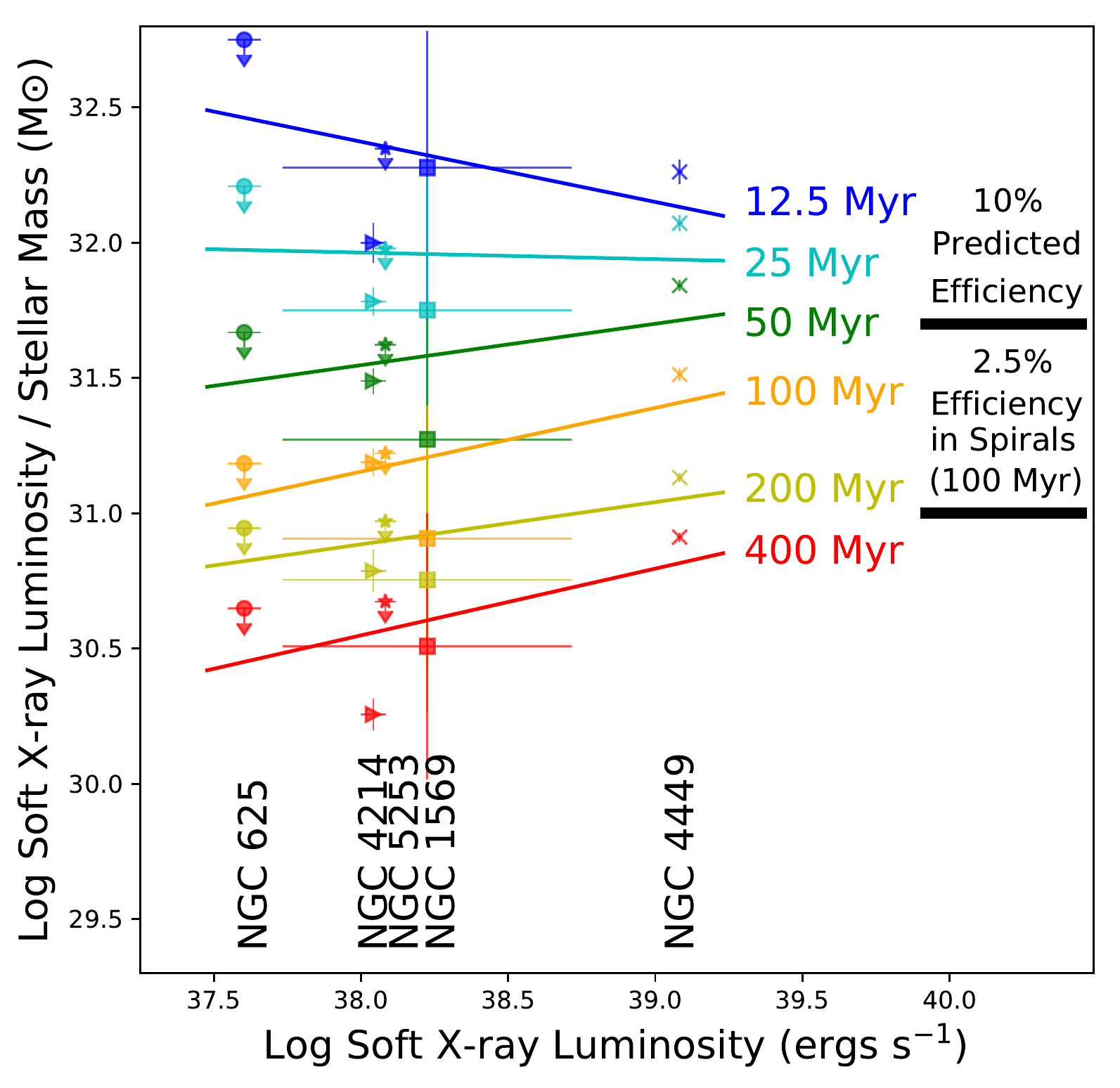} 
\caption{Diffuse, soft X-ray luminosity corrected for extinction compared with the same X-ray luminosity normalized by the stellar mass formed over increasing periods of time from the SFHs (i.e., ranging from $0-12.5$ Myr to $0-400$ Myr) for the 5 galaxies with detected soft X-ray emission. The change in slope over the most recent 100 Myr is driven primarily by the post-starburst galaxy NGC~625 and NGC~5253, suggesting an upper limit for longevity timescale for X-ray emission of 25 Myr. Also noted are the equivalent 10\% efficiency for converting mechanical energy from star formation to soft X-ray emission from simulations of low-mass galaxies \citep{Strickland2000a} and the 2.5\% efficiency from measurements of spiral galaxies over a similar energy range \citep{Mineo2012}.}
\label{fig:xray_stellar_mass}
\end{figure}

\subsection{Longevity of an Outflow}
Once starburst activity has ended, it is an open question how long an outflow persists. One of the few constraints come from theoretically derived cooling timescales of hot gas. Using spectral models of X-ray emission, cooling timescales have been estimated to range from $\sim20-400$ Myr \citep[e.g.,][]{Summers2003, Hartwell2004}. These values are necessarily uncertain and likely overestimate cooling times as they assume spherical symmetry of X-ray emission and filling factors of unity. An additional constraint was set by numerical simulations of galactic winds in dwarf galaxies which predict a timescale of $\sim10-15$ Myr \citep{Strickland2000a}. 

An ideal observational experiment would be to measure the diffuse X-ray emission in a large sample of galaxies that are both bursting and post-burst with star formation activity declining over different timescales. The expectation is that the present-day diffuse soft X-ray luminosity relative to past star formation activity (i.e., Lxray$_{soft}$/SFR(t)) will be approximately constant for some period of time even after the starburst declines. Thus, comparing the Lxray$_{soft}$ to the amount of stellar mass formed over varying time periods and identifying when the time frame over which Lxray$_{soft}$/M$_*$(t) is constant can constrain the longevity of the wind. Our current sample includes galaxies that are currently bursting, one galaxy with declining SFRs over the past 100 Myr, and one post-starburst system. While only a small sample, we explore whether we can detect the timescale of winds by comparing the star formation activity over different time bins from the SFHs.

In Figure~\ref{fig:xray_stellar_mass} we present a comparison of the measured soft X-ray luminosities with the soft X-ray luminosities normalized by the stellar mass formed in logarithmically spaced time bins of $0-12.5, 0-25, 0-50, 0-100, 0-200, 0-400$ Myr from the SFHs in \citet{McQuinn2010b}. Each galaxy is plotted with a unique symbol and each line represents the least-squares fit for a given time bin across the sample. Note that the X-ray luminosity for each galaxy is the currently measured diffuse flux while the stellar mass measurements give a historical record of the star formation activity over the past 400 Myr. Thus, the y-axis probes current wind conditions relative to past events. The stellar mass values for NGC~4214 and NGC~625 are lower limits as the $HST$ footprints did not encompass the full stellar discs of the galaxies.

Generally, the diffuse soft X-ray luminosity is expected to scale with the number of stars formed. The study of spiral galaxies from $Chandra$ data suggest a linear relationship \citep{Mineo2012}, in agreement with simulations \cite[e.g.,][]{Bustard2016, Meiksin2016, vandeVoort2016}, although there is still some range in X-ray emission predicted at a given SFR. In Figure~\ref{fig:xray_stellar_mass}, a linear relationship between X-ray luminosity and stellar mass corresponds to a flat slope (seen at 25 Myr), whereas the steeper relation predicted by simulations corresponds to a positive slope (seen for timescales 50 Myr and greater). The shortest timescale shown is 12.5 Myr, comparable to the formation timescale of a wind, has higher than expected X-ray luminosity, contradicting results from the studies of spiral galaxies and simulations, and supporting our interpretation that this short star formation timescale does not correlate directly with wind properties.

While the conclusion is limited by the small number of points, the changing slope over the last 100 Myr suggests an upper limit of 25 Myr for a wind timescale. This is only slightly longer than the dynamical timescale for the development of an outflow, but note that these timescales are measuring different events. Outflow formation includes the expansion of lower density regions powered by SNe through the ISM, whereas we are measuring the longevity of a post-starburst outflow. The total duration of a wind is then the formation timescale ($\sim10-25$ Myr), plus the ongoing outflow phase (of order the duration of the starbursts of few 100 Myr), plus the longevity timescale (25 Myr), before the X-ray surface brightness falls below current detection limits due to cooling and lower column densities.

%
\begin{table*}
\begin{center}
\begin{tabular}{l c | cc | cc | cc | cc}
\hline
\hline 	
			& Angular	& \multicolumn{2}{c}{10 Myr}			& \multicolumn{2}{c}{25 Myr}			& \multicolumn{2}{c}{100 Myr}				& \multicolumn{2}{c}{Burst Duration} \\
		  	&  X-ray 	&  $\rm{L_{mech}}$	& $\rm{L_{Xray}}$/	& $\rm{L_{mech}}$	& $\rm{L_{Xray}}$/	& $\rm{L_{mech}}$	& $\rm{L_{Xray}}/$ 	&  $\rm{L_{mech}}$	& $\rm{L_{Xray}}$/	\\
Galaxy	  	&  Extent 	&  $\times$10$^{38}$& $\rm{L_{mech}}$	& $\times$10$^{38}$ & $\rm{L_{mech}}$	& $\times10^{38}$	& $\rm{L_{mech}}$ 	&  $\times$10$^{38}$& $\rm{L_{mech}}$	\\
 			&(2$\sigma$) & (ergs s$^{-1}$)	& 				& (ergs s$^{-1}$)	& 				& (ergs s$^{-1}$)	& 				& (ergs s$^{-1}$)	& 			\\
(1) & (2) & (3) & (4) & (5) & (6) & (7) & (8) & (9) & (10) \\
\hline
DDO~165         	& ...			& 0.6			& ...			& 3.5		& ...		& 9.2 	& ...		& 77		& ...\\ 
NGC~625        	& 1.6\arcmin	& $>$0.2		& $<$183\%	& 1.3		& $<$32\%& 20. 	& $<$3\% 	& $>$71	& $<$ 0.6\% \\
NGC~1569      	& 4.8\arcmin	& 3.9			& 42\%		& 15		& 11\%	& 101	& 2\%	& 280	& 0.6\% \\
NGC~4214     	& 6.8\arcmin	& $>$1.8		& $<$69\%	& $>$3.5	& $<$35\% &$>$20. & $<$2\%	& $>$200	& $<$0.6\% \\
NGC~4449     	& 4.9\arcmin	& 61			& 20\%		& 55		& 22\%	&200 	& 6\%	& 1100	& 1.1\% \\
NGC~5253     	& 4.0\arcmin	& 8.8			& 12\%		& 9.9		& 11\%	& 30 		& 3\%	& 560	& 0.2\% \\
\hline
\end{tabular}
\end{center}
\caption{Star Formation Energetics and X-ray Luminosities. The ratio of X-ray emission to star formation energetics is higher than predictions on short timescales (25 Myr), while significantly lower over burst timescales. Col. 2 lists the largest angular extent of the X-ray emission at the 2$\sigma$ contour levels shown in Figures~\ref{fig:ddo165_xraycontour}$-$\ref{fig:ngc5253_xraycontour}. Cols. 3, 5, 7, and 9. Mechanical luminosity based on star formation over the past 10 Myr, 25 Myr, 100 Myr, and the duration of the starbursts calculated using the SFHs and STARBURS99 model. Cols. 4, 6, 8, and 10. Ratio of the measured X-ray to mechanical luminosities. Mechanical luminosities for NGC~625 and NGC~4214 are lower limits as the $HST$ field of view does not cover the full optical discs of the galaxies.}
\label{tab:xray_mech}
\end{table*}

Recent simulations of stellar-feedback driven galactic winds have predicted the diffuse, soft (0.5-2.0 keV) X-ray surface brightness as a function of time and galactic radius \citep{vandeVoort2016}. These results indicate that an outflow produced by a discrete star formation episode propagates slowly in the CGM and reaches a galaxy's virial radius in $\sim500$ Myr, with average X-ray surface brightnesses in the inner 0.05 virial radius (i.e., $\sim$ 5 kpc) declining from $10^{-17}$ erg s$^{-1}$ cm$^{-2}$ arcsec$^{-2}$ to 10$^{-23}$ erg s$^{-1}$ cm$^{-2}$ arcsec$^{-2}$. Our average X-ray surface brightnesses range from $\sim10^{-17}$ to $\sim 10^{-19}$ erg s$^{-1}$ cm$^{-2}$ arcsec$^{-2}$ for radii within 5 kpc. These values are at the upper end of the predicted range which corresponds to the initial $\sim60$ Myr time interval in the simulations and provides a consistency check on the 25 Myr longevity timescale of the wind from our analysis. 

\subsection{Wind Efficiencies}
A complementary approach to our comparison of soft X-ray luminosity to stellar mass is to compare the soft X-ray luminosity with the energy produced by star formation. This is sometimes referred to as the efficiency of converting the mechanical energy of stellar winds and SNe to the thermal energy of hot winds. We include this calculation as it allows us to compare our results more directly with previous studies, but note that it also introduces an additional assumption from modelling the mechanical energy of stellar feedback rather than the more directly observable quantity of stellar mass used in Figure~\ref{fig:xray_stellar_mass}. 

Previous studies report wind efficiencies below 10\%. In numerical simulations, \citet{Strickland2000a} found that the modelled soft X-ray luminosities represented $\ltsimeq10$\% of the energy injected from star formation in low-mass galaxies; a significant fraction of the remaining energy is kinetic with a smaller fraction from thermal emission in higher temperature gas. This study assumed short star formation timescales (ranging from an instantaneous burst to 10 Myr) and 100\% of the mechanical power from stellar winds and SNe is available to power the wind (i.e., radiative losses are negligible). 

In a study of spiral galaxies from $Chandra$ deep fields, a 5\% efficiency is implied for converting the mechanical energy from star formation rates averaged over 100 Myr timescales to bolometric X-ray luminosities. This efficiency corresponds to a value of 2.5\% when considering the X-ray luminosity over the energy range of 0.3-10.0 keV \citep{Mineo2012}, which is comparable to the efficiency over a soft X-ray band found using the PIMMS software.

We chose four timescales from the SFHs to calculate the wind efficiency, namely a short time bin (i.e., 12.5 Myr which is comparable to the 10 Myr used in simulations by \citet{Strickland2000a}), a 25 Myr timescale corresponding to the upper limit for our estimated wind longevity, a 100 Myr timescale comparable to the timescale in the $Chandra$ study of spiral galaxies \citep{Mineo2012}, and the full duration timescale of the bursts. First, we model the energy input from stellar winds and SNe from the SFRs over the last 12.5 Myr using the STARBURST99 code \citep{Leitherer1999} and assuming a Salpeter initial mass function \citep{Salpeter1955}. Second, we repeat the modelling over 25 Myr , 100 Myr, and the lifetime of the burst events (which is a timescale not probed in our comparison with stellar mass in Figure~\ref{fig:xray_stellar_mass}). 

Table~\ref{tab:xray_mech} lists the modelled mechanical luminosities ($\rm{L_{mech}}$) of the stellar feedback from stellar winds and SNe over the four timescales, along with the ratios of the measured soft X-ray luminosities to the mechanical luminosities. For the 12.5 Myr timescale, the thermal X-ray emission relative to mechanical energy of star formation is between 12-42\% for galaxies whose disks are fully sampled in the $HST$ field of view; we set upper limits of  69\% (NGC~4214) and 183\% (NGC~625) for the two galaxies with larger angular sizes. Note, however, that in the case of NGC~625 the upper limit is only marginally driven by the smaller $HST$ areal coverage. The more dominant effect is the declining star formation rate in this post-starburst galaxy and smaller amount of stellar mass formed in the past 12.5 Myr. For the 25 Myr timescales, the ratios of $\rm{L_{xray} / L_{mech}}$ are between 11-22\% for galaxies whose stellar discs are within the $HST$ field of view, and upper limit of 35\% and 32\% for the two remaining systems. For the 100 Myr timescales, the ratios of $\rm{L_{xray} / L_{mech}}$ are between 1-6\% for galaxies whose stellar discs are within the $HST$ field of view, and upper limit of 6\% and 3\% for the two remaining systems. Finally, for the longer burst timescales, the ratios of $\rm{L_{xray} / L_{mech}}$ are all $\ltsimeq1$\%. 

In Figure~\ref{fig:xray_stellar_mass}, we show the equivalent value of a 10\% efficiency representing the upper bound from the above studies. Our measured X-ray luminosities relative to star formation activity are higher than the predicted values for nearly all galaxies and across all timescales considered, indicating a higher efficiency for the transfer of energy. Our calculated efficiencies are up to $4\times$ higher than predicted by the simulations of \citet{Strickland2000a} over the shortest timescale. For the 25 Myr timescale, we calculate an average efficiency of 15\%, excluding the upper limit from NGC~4214 and NGC~625, which is a closer match to the predictions. 

In Figure~\ref{fig:xray_stellar_mass}, we also show the equivalent value of a 2.5\% efficiency found for 100 Myr timescales in spiral galaxies from \citet{Mineo2012}. Note that the efficiencies from \citet{Mineo2012} were estimated using X-ray luminosities across a slightly difference energy range of 0.5-2.0 keV that, corrected for unresolved point sources, lowers the measured luminosities (and efficiencies) by 37\% . Based on their spectral analysis, there is a smaller $\sim10$\% contribution of unresolved point sources in our chosen energy range of 0.35 - 1.0 keV. Thus, our calculated efficiencies are approximately consistent with those estimated for spiral galaxies within the uncertainties. 

Finally, we note that over the longer timescales of the burst durations, the low ($\ltsimeq$1\%) efficiencies suggest that much of the mechanical energy input into the ISM  is ``missing'' and has either been radiatively cooled, transported out of the galaxy with surface brightness levels below our detection limit, or both. This indicates that the outflows in these systems must have been on-going for a significant fraction of the burst durations. It is interesting to note that wind recycling times (i.e., the time for gas to be re-accreted to the host galaxy from the CGM) are predicted to be of order 300 Myr from simulations  \citep{Angles-Alcazar2016} $-$ comparable to the burst durations in some of our galaxies $-$ but it is unclear how this relates to temporally extended outflows  versus outflows powered by discrete star formation events.

\subsection{Re-Examination of Individual Outflows}\label{sec:reexamine}
Given the additional constraints provided by the SFHs, we re-examine the galaxies and outflows in our sample, taking into account previous measurements and results. 

For NGC~4449, \citet{Summers2003} calculated that the outflow would take 200 Myr to reach the edge of the large 40 kpc gas halo (in projection) based on estimated gas velocities of $\sim220-280$ km s$^{-1}$. These authors conclude the outflow will stall before reaching the edge of the gas halo based on the energy ejection from star formation on short timescales $<10$ Myr (i.e., the lifetime of an OB association). However, from the SFHs, the starburst  in NGC~4449 has been on-going for the past $450\pm50$ Myr.  From Figure~\ref{fig:ngc4449_xraycontour}, the hot gas traced by the soft X-ray emission extends past the \HI\ along the NW-SE axis suggesting that material is already being expelled out of the main \HI\ disc  via these pathways. Because of the very low implied amount of the energy from stellar feedback present in current thermal X-ray emission ($<$1\%), the outflow in NGC~4449 must have been on-going for a significant fraction of the burst lifetime. Depending on the 3-D distribution of the extended \HI, this material may not have been expelled into the IGM, but may instead reside at extended radii around the galaxy and/or recycled back into the galaxy.

For NGC~5253, the detected outflow surrounds much of the central star forming region and is strongest in the south. \citet{Marlowe1995} measure velocities of the \ha\ detected shells of 35 km s$^{-1}$ with physical extents of $\sim1$ kpc. \citet{Summers2004} find the extent of the X-ray emission to be somewhat smaller ($\sim0.5$ kpc). However, using combined {\it Chandra} observations, we find a more extended region in the south of the galaxy that reaches $\sim2.4$ kpc in projection which cannot be explained by dynamical arguments and short star formation timescales. These inconsistencies are alleviated when the longer star formation timescales from the SFHs are considered. Furthermore, the SFR in NGC~5253 has declined by nearly a factor of 6 over the duration of the burst. The declining SFRs and small-scale outflow in NGC~5253 indicate that we are seeing the end of a wind phase (see also Figure~\ref{fig:xray_stellar_mass}), rather than just the beginning of an outflow as has been previously suggested \citep{Summers2004}.

For NGC~1569, the dynamical age of the outflow was previously estimated to be 10 Myr \citep{Heckman1995, Martin1998}. Using this timescale, \citet{Martin2002} estimate the amount of oxygen that could be in the wind of NGC~1569 and compare this to the amount of metals produced by star formation over a 10-20 Myr timescale assuming a nucleosynthesis yield. Based on their best-fitting model of a solar metallicity in the wind, and to maintain consistency with the current gas-phase oxygen abundance in the galaxy, these authors report nearly all metals formed in the recent star formation activity have been expelled in the wind. However, as stated in \citet{Martin2002}, this requires that the present-day gas-phase oxygen content in the galaxy was created in previous episodes of star formation that must have been quiescent (i.e., not energetic enough to drive an outflow). Based on the CMDs and the derived SFHs, there has been significant star formation in the last few 100 Myr in NGC~1569, with the peak star formation occurring 65 Myr. This suggests that the wind has been ongoing. Furthermore, the best-fitting model of a solar oxygen abundance in the hot wind does not take into account the uncertainties; a nearly equal $\chi^2$ is found for 0.25 \zsun, as well as for super-solar abundances for NGC~1569 \citep[][see their Table~5]{Martin2002}. The amount of metals ejected by the wind in NGC~1569 remains an open question.

\section{Conclusions}\label{sec:conclusions}
We have examined the diffuse soft X-ray emission from {\it Chandra} observations in six nearby starburst dwarf galaxies from the STARBIRDS sample \citep{McQuinn2015b} and compared the X-ray properties with the neutral gas distributions, SFHs quantified from resolved stellar populations, and UV emission. Similar to the previous studies, we detect an outflow in NGC~1569, traced by diffuse, soft X-ray emitting gas,  and extended X-ray emission in the face-on galaxy NGC~4214. We also detect more clearly defined outflows from NGC~4449 and NGC~5253 than in previous studies by combining data from multiple observing and using a slightly broader energy range to define our soft X-ray energy band. We find the end of an outflow phase in NGC~625. No outflow was detected in DDO~165; this galaxy has likely experienced a large-scale blow-out of the ISM in the last $\ltsimeq100$ Myr \citep{Cannon2011a, Cannon2011b}. 

The starburst activity has been declining over the most recent 100 Myr in two galaxies (NGC~625, NGC~5253), allowing us to constrain the longevity of the winds to have an upper limit of 25 Myr. While based on a small sample, this is the first empirical measurement of a wind timescale. 

Comparing the X-ray luminosities to the mechanical energy injected by stellar feedback over different timescales, we find the diffuse soft X-ray emission accounts for $11-22$\% of the energy from star formation deposited over short 25 Myr timescales (with upper limits from two systems of 35\% and 32\%), but represents $\ltsimeq1$\% of the energy from the full starburst events based on modelling from the SFHs. These ratios bracket the starburst event and indicate that the outflows have been sustained for a significant fraction of the burst duration. Such temporally extended outflows suggest that expelled material may reach farther distances before cooling, increasing the amount of material that will reach the CGM or IGM and potentially lowering the recycling fraction of gas back to the ISM. Over the wind timescale of 25 Myr, the average wind efficiency is 16\%. This can be compared with $<10$\% predicted from numerical simulations on 10 Myr timescales \citep{Strickland2000a} and 2.5\% estimated for spiral galaxies on 100 Myr timescales \citep{Mineo2012}. 

It is also interesting to note that none of the outflows have a biconical morphology often assumed for winds and seen in, for example, the extreme system M82. The strongest outflow detected is in NGC~1569, where expelled material extends across the length of the stellar disc, above and below the minor axis. Previously described as a bipolar outflow \citep{Martin2002}, the morphology is amorphous and agrees with the chaotic distribution of hot gas produced in simulations \citep[e.g.,][]{vandeVoort2016}. 

Hydrodynamical simulations that are now resolving low-mass galaxies show that stellar feedback may not only drive many of the observed properties of dwarfs, but may also be the key to distinguishing between different cosmologies. Observationally, our empirical measurements and constraints for hot winds for the simulations come from a small number of low-mass galaxies, including the six galaxies studied here. Given the limited observations, it is still unclear whether outflows in low-mass galaxies occur as ubiquitously as predicted, or whether their characteristics agree with predictions over a range of galaxy masses. 

In this work we focused on the mechanical energy of star formation converted into X-ray emission in the hot-phase wind. In a companion paper for STARBIRDS, we address the mass-loading factors of winds measured in the warm, ionized phase of the ISM, where the majority of gas mass expelled from galaxies is expected (McQuinn et al.\ in prep).

\section*{Acknowledgments}
This study was funded by the Smithsonian Astrophysical Observatory Award AR2-13015X under NASA contract NAS 8-03060. This research has made use of NASA's Astrophysics Data System, the ``Aladin sky atlas'' developed at CDS, Strasbourg Observatory, France \citep{Boch2014, Bonnarel2000}, the {{\em ACIS Extract}} software package \citep{Broos2010}, and the NASA/IPAC Extragalactic Database (NED), which is operated by the Jet Propulsion Laboratory, California Institute of Technology, under contract with the National Aeronautics and Space Administration. We acknowledge the usage of the HyperLeda database (http://leda.univ-lyon1.fr). We would like to thank the referee for helpful comments and suggestions that have improved the manuscript and Nick Lee at the Chandra X-ray Center for providing valuable expertise on the data reduction process. 

\bibliographystyle{mnras}
\bibliography{../../bibliography}

\begin{thebibliography}{}
\makeatletter
\relax
\def\mn@urlcharsother{\let\do\@makeother \do\$\do\&\do\#\do\^\do\_\do\%\do\~}
\def\mn@doi{\begingroup\mn@urlcharsother \@ifnextchar [ {\mn@doi@}
  {\mn@doi@[]}}
\def\mn@doi@[#1]#2{\def\@tempa{#1}\ifx\@tempa\@empty \href
  {http://dx.doi.org/#2} {doi:#2}\else \href {http://dx.doi.org/#2} {#1}\fi
  \endgroup}
\def\mn@eprint#1#2{\mn@eprint@#1:#2::\@nil}
\def\mn@eprint@arXiv#1{\href {http://arxiv.org/abs/#1} {{\tt arXiv:#1}}}
\def\mn@eprint@dblp#1{\href {http://dblp.uni-trier.de/rec/bibtex/#1.xml}
  {dblp:#1}}
\def\mn@eprint@#1:#2:#3:#4\@nil{\def\@tempa {#1}\def\@tempb {#2}\def\@tempc
  {#3}\ifx \@tempc \@empty \let \@tempc \@tempb \let \@tempb \@tempa \fi \ifx
  \@tempb \@empty \def\@tempb {arXiv}\fi \@ifundefined
  {mn@eprint@\@tempb}{\@tempb:\@tempc}{\expandafter \expandafter \csname
  mn@eprint@\@tempb\endcsname \expandafter{\@tempc}}}

\bibitem[\protect\citeauthoryear{{Angl{\'e}s-Alc{\'a}zar},
  {Faucher-Gigu{\`e}re}, {Kere{\v s}}, {Hopkins}, {Quataert}  \&
  {Murray}}{{Angl{\'e}s-Alc{\'a}zar} et~al.}{2016}]{Angles-Alcazar2016}
{Angl{\'e}s-Alc{\'a}zar} D.,  {Faucher-Gigu{\`e}re} C.-A.,  {Kere{\v s}} D.,
  {Hopkins} P.~F.,  {Quataert} E.,   {Murray} N.,  2016, preprint, \href
  {http://adsabs.harvard.edu/abs/2016arXiv161008523A} {} (\mn@eprint {arXiv}
  {1610.08523})

\bibitem[\protect\citeauthoryear{{Berg} et~al.,}{{Berg}
  et~al.}{2012}]{Berg2012}
{Berg} D.~A.,  et~al., 2012, \mn@doi [\apj] {10.1088/0004-637X/754/2/98}, \href
  {http://adsabs.harvard.edu/abs/2012ApJ...754...98B} {754, 98}

\bibitem[\protect\citeauthoryear{{Binder} et~al.,}{{Binder}
  et~al.}{2012}]{Binder2012}
{Binder} B.,  et~al., 2012, \mn@doi [\apj] {10.1088/0004-637X/758/1/15}, \href
  {http://adsabs.harvard.edu/abs/2012ApJ...758...15B} {758, 15}

\bibitem[\protect\citeauthoryear{{Bland} \& {Tully}}{{Bland} \&
  {Tully}}{1988}]{Bland1988}
{Bland} J.,  {Tully} B.,  1988, \mn@doi [\nat] {10.1038/334043a0}, \href
  {http://adsabs.harvard.edu/abs/1988Natur.334...43B} {334, 43}

\bibitem[\protect\citeauthoryear{{Boch} \& {Fernique}}{{Boch} \&
  {Fernique}}{2014}]{Boch2014}
{Boch} T.,  {Fernique} P.,  2014, in {Manset} N.,  {Forshay} P.,  eds,
  Astronomical Society of the Pacific Conference Series Vol. 485, Astronomical
  Data Analysis Software and Systems XXIII. p.~277

\bibitem[\protect\citeauthoryear{{Bonnarel} et~al.,}{{Bonnarel}
  et~al.}{2000}]{Bonnarel2000}
{Bonnarel} F.,  et~al., 2000, \mn@doi [\aaps] {10.1051/aas:2000331}, \href
  {http://adsabs.harvard.edu/abs/2000A%26AS..143...33B} {143, 33}

\bibitem[\protect\citeauthoryear{{Boylan-Kolchin}, {Bullock}  \&
  {Kaplinghat}}{{Boylan-Kolchin} et~al.}{2011}]{Boylan-Kolchin2011}
{Boylan-Kolchin} M.,  {Bullock} J.~S.,   {Kaplinghat} M.,  2011, \mn@doi
  [\mnras] {10.1111/j.1745-3933.2011.01074.x}, \href
  {http://adsabs.harvard.edu/abs/2011MNRAS.415L..40B} {415, L40}

\bibitem[\protect\citeauthoryear{{Broos}, {Townsley}, {Feigelson}, {Getman},
  {Bauer}  \& {Garmire}}{{Broos} et~al.}{2010}]{Broos2010}
{Broos} P.~S.,  {Townsley} L.~K.,  {Feigelson} E.~D.,  {Getman} K.~V.,  {Bauer}
  F.~E.,   {Garmire} G.~P.,  2010, \mn@doi [\apj]
  {10.1088/0004-637X/714/2/1582}, \href
  {http://adsabs.harvard.edu/abs/2010ApJ...714.1582B} {714, 1582}

\bibitem[\protect\citeauthoryear{{Broos} et~al.,}{{Broos}
  et~al.}{2011}]{Broos2011}
{Broos} P.~S.,  et~al., 2011, \mn@doi [\apjs] {10.1088/0067-0049/194/1/2},
  \href {http://adsabs.harvard.edu/abs/2011ApJS..194....2B} {194, 2}

\bibitem[\protect\citeauthoryear{{Bustard}, {Zweibel}  \& {D'Onghia}}{{Bustard}
  et~al.}{2016}]{Bustard2016}
{Bustard} C.,  {Zweibel} E.~G.,   {D'Onghia} E.,  2016, \mn@doi [\apj]
  {10.3847/0004-637X/819/1/29}, \href
  {http://adsabs.harvard.edu/abs/2016ApJ...819...29B} {819, 29}

\bibitem[\protect\citeauthoryear{{Cannon}, {McClure-Griffiths}, {Skillman}  \&
  {C{\^o}t{\'e}}}{{Cannon} et~al.}{2004}]{Cannon2004}
{Cannon} J.~M.,  {McClure-Griffiths} N.~M.,  {Skillman} E.~D.,   {C{\^o}t{\'e}}
  S.,  2004, \mn@doi [\apj] {10.1086/383408}, \href
  {http://adsabs.harvard.edu/abs/2004ApJ...607..274C} {607, 274}

\bibitem[\protect\citeauthoryear{{Cannon} et~al.,}{{Cannon}
  et~al.}{2011a}]{Cannon2011a}
{Cannon} J.~M.,  et~al., 2011a, \mn@doi [\apj] {10.1088/0004-637X/735/1/35},
  \href {http://adsabs.harvard.edu/abs/2011ApJ...735...35C} {735, 35}

\bibitem[\protect\citeauthoryear{{Cannon} et~al.,}{{Cannon}
  et~al.}{2011b}]{Cannon2011b}
{Cannon} J.~M.,  et~al., 2011b, \mn@doi [\apj] {10.1088/0004-637X/735/1/36},
  \href {http://adsabs.harvard.edu/abs/2011ApJ...735...36C} {735, 36}

\bibitem[\protect\citeauthoryear{{Dalcanton}}{{Dalcanton}}{2007}]{Dalcanton2007}
{Dalcanton} J.~J.,  2007, \mn@doi [\apj] {10.1086/508913}, \href
  {http://adsabs.harvard.edu/abs/2007ApJ...658..941D} {658, 941}

\bibitem[\protect\citeauthoryear{{Dickey} \& {Lockman}}{{Dickey} \&
  {Lockman}}{1990}]{Dickey1990}
{Dickey} J.~M.,  {Lockman} F.~J.,  1990, \mn@doi [\araa]
  {10.1146/annurev.aa.28.090190.001243}, \href
  {http://adsabs.harvard.edu/abs/1990ARA%26A..28..215D} {28, 215}

\bibitem[\protect\citeauthoryear{{Dolphin}}{{Dolphin}}{2002}]{Dolphin2002a}
{Dolphin} A.~E.,  2002, \mn@doi [\mnras] {10.1046/j.1365-8711.2002.05271.x},
  \href {http://adsabs.harvard.edu/abs/2002MNRAS.332...91D} {332, 91}

\bibitem[\protect\citeauthoryear{{Finlator} \& {Dav{\'e}}}{{Finlator} \&
  {Dav{\'e}}}{2008}]{Finlator2008}
{Finlator} K.,  {Dav{\'e}} R.,  2008, \mn@doi [\mnras]
  {10.1111/j.1365-2966.2008.12991.x}, \href
  {http://adsabs.harvard.edu/abs/2008MNRAS.385.2181F} {385, 2181}

\bibitem[\protect\citeauthoryear{{Freeman}, {Doe}  \&
  {Siemiginowska}}{{Freeman} et~al.}{2001}]{Freeman2001}
{Freeman} P.,  {Doe} S.,   {Siemiginowska} A.,  2001, in {Starck} J.-L.,
  {Murtagh} F.~D.,  eds,  \procspie Vol. 4477, Astronomical Data Analysis. pp
  76--87 (\mn@eprint {} {astro-ph/0108426}), \mn@doi{10.1117/12.447161}

\bibitem[\protect\citeauthoryear{{Garmire}, {Bautz}, {Ford}, {Nousek}  \&
  {Ricker}}{{Garmire} et~al.}{2003}]{Garmire2003}
{Garmire} G.~P.,  {Bautz} M.~W.,  {Ford} P.~G.,  {Nousek} J.~A.,   {Ricker} Jr.
  G.~R.,  2003, in {Truemper} J.~E.,  {Tananbaum} H.~D.,  eds,  \procspie Vol.
  4851, X-Ray and Gamma-Ray Telescopes and Instruments for Astronomy.. pp
  28--44, \mn@doi{10.1117/12.461599}

\bibitem[\protect\citeauthoryear{{Garnett}}{{Garnett}}{2002}]{Garnett2002}
{Garnett} D.~R.,  2002, \mn@doi [\apj] {10.1086/344301}, \href
  {http://adsabs.harvard.edu/abs/2002ApJ...581.1019G} {581, 1019}

\bibitem[\protect\citeauthoryear{{Grocholski} et~al.,}{{Grocholski}
  et~al.}{2008}]{Grocholski2008}
{Grocholski} A.~J.,  et~al., 2008, \mn@doi [\apjl] {10.1086/592949}, \href
  {http://adsabs.harvard.edu/abs/2008ApJ...686L..79G} {686, L79}

\bibitem[\protect\citeauthoryear{{Hartwell}, {Stevens}, {Strickland}, {Heckman}
   \& {Summers}}{{Hartwell} et~al.}{2004}]{Hartwell2004}
{Hartwell} J.~M.,  {Stevens} I.~R.,  {Strickland} D.~K.,  {Heckman} T.~M.,
  {Summers} L.~K.,  2004, \mn@doi [\mnras] {10.1111/j.1365-2966.2004.07375.x},
  \href {http://adsabs.harvard.edu/abs/2004MNRAS.348..406H} {348, 406}

\bibitem[\protect\citeauthoryear{{Heckman}, {Dahlem}, {Lehnert}, {Fabbiano},
  {Gilmore}  \& {Waller}}{{Heckman} et~al.}{1995}]{Heckman1995}
{Heckman} T.~M.,  {Dahlem} M.,  {Lehnert} M.~D.,  {Fabbiano} G.,  {Gilmore} D.,
    {Waller} W.~H.,  1995, \mn@doi [\apj] {10.1086/175944}, \href
  {http://adsabs.harvard.edu/abs/1995ApJ...448...98H} {448, 98}

\bibitem[\protect\citeauthoryear{{Hopkins}, {Quataert}  \& {Murray}}{{Hopkins}
  et~al.}{2012}]{Hopkins2012}
{Hopkins} P.~F.,  {Quataert} E.,   {Murray} N.,  2012, \mn@doi [\mnras]
  {10.1111/j.1365-2966.2012.20593.x}, \href
  {http://adsabs.harvard.edu/abs/2012MNRAS.421.3522H} {421, 3522}

\bibitem[\protect\citeauthoryear{{Hunter}, {Wilcots}, {van Woerden},
  {Gallagher}  \& {Kohle}}{{Hunter} et~al.}{1998}]{Hunter1998}
{Hunter} D.~A.,  {Wilcots} E.~M.,  {van Woerden} H.,  {Gallagher} J.~S.,
  {Kohle} S.,  1998, \mn@doi [\apjl] {10.1086/311213}, \href
  {http://adsabs.harvard.edu/abs/1998ApJ...495L..47H} {495, L47}

\bibitem[\protect\citeauthoryear{{Jacobs}, {Rizzi}, {Tully}, {Shaya}, {Makarov}
   \& {Makarova}}{{Jacobs} et~al.}{2009}]{Jacobs2009}
{Jacobs} B.~A.,  {Rizzi} L.,  {Tully} R.~B.,  {Shaya} E.~J.,  {Makarov} D.~I.,
   {Makarova} L.,  2009, \mn@doi [\aj] {10.1088/0004-6256/138/2/332}, \href
  {http://adsabs.harvard.edu/abs/2009AJ....138..332J} {138, 332}

\bibitem[\protect\citeauthoryear{{Kaastra} \& {Mewe}}{{Kaastra} \&
  {Mewe}}{1993}]{Kaastra1993}
{Kaastra} J.~S.,  {Mewe} R.,  1993, \aaps, \href
  {http://adsabs.harvard.edu/abs/1993A%26AS...97..443K} {97, 443}

\bibitem[\protect\citeauthoryear{{Kalberla}, {Burton}, {Hartmann}, {Arnal},
  {Bajaja}, {Morras}  \& {P{\"o}ppel}}{{Kalberla} et~al.}{2005}]{Kalberla2005}
{Kalberla} P.~M.~W.,  {Burton} W.~B.,  {Hartmann} D.,  {Arnal} E.~M.,  {Bajaja}
  E.,  {Morras} R.,   {P{\"o}ppel} W.~G.~L.,  2005, \mn@doi [\aap]
  {10.1051/0004-6361:20041864}, \href
  {http://adsabs.harvard.edu/abs/2005A%26A...440..775K} {440, 775}

\bibitem[\protect\citeauthoryear{{Kennicutt}, {Tamblyn}  \&
  {Congdon}}{{Kennicutt} et~al.}{1994}]{Kennicutt1994}
{Kennicutt} Jr. R.~C.,  {Tamblyn} P.,   {Congdon} C.~E.,  1994, \mn@doi [\apj]
  {10.1086/174790}, \href {http://adsabs.harvard.edu/abs/1994ApJ...435...22K}
  {435, 22}

\bibitem[\protect\citeauthoryear{{Kim}, {Ostriker}  \& {Raileanu}}{{Kim}
  et~al.}{2017}]{Kim2017}
{Kim} C.-G.,  {Ostriker} E.~C.,   {Raileanu} R.,  2017, \mn@doi [\apj]
  {10.3847/1538-4357/834/1/25}, \href
  {http://adsabs.harvard.edu/abs/2017ApJ...834...25K} {834, 25}

\bibitem[\protect\citeauthoryear{{Klypin}, {Kravtsov}, {Valenzuela}  \&
  {Prada}}{{Klypin} et~al.}{1999}]{Klypin1999}
{Klypin} A.,  {Kravtsov} A.~V.,  {Valenzuela} O.,   {Prada} F.,  1999, \mn@doi
  [\apj] {10.1086/307643}, \href
  {http://adsabs.harvard.edu/abs/1999ApJ...522...82K} {522, 82}

\bibitem[\protect\citeauthoryear{{Kobulnicky} \& {Skillman}}{{Kobulnicky} \&
  {Skillman}}{1995}]{Kobulnicky1995}
{Kobulnicky} H.~A.,  {Skillman} E.~D.,  1995, \mn@doi [\apjl] {10.1086/309791},
  \href {http://adsabs.harvard.edu/abs/1995ApJ...454L.121K} {454, L121}

\bibitem[\protect\citeauthoryear{{Kobulnicky} \& {Skillman}}{{Kobulnicky} \&
  {Skillman}}{2008}]{Kobulnicky2008}
{Kobulnicky} H.~A.,  {Skillman} E.~D.,  2008, \mn@doi [\aj]
  {10.1088/0004-6256/135/2/527}, \href
  {http://adsabs.harvard.edu/abs/2008AJ....135..527K} {135, 527}

\bibitem[\protect\citeauthoryear{{Lee}, {Skillman}  \& {Venn}}{{Lee}
  et~al.}{2006}]{Lee2006}
{Lee} H.,  {Skillman} E.~D.,   {Venn} K.~A.,  2006, \mn@doi [\apj]
  {10.1086/500568}, \href {http://adsabs.harvard.edu/abs/2006ApJ...642..813L}
  {642, 813}

\bibitem[\protect\citeauthoryear{{Leitherer} et~al.,}{{Leitherer}
  et~al.}{1999}]{Leitherer1999}
{Leitherer} C.,  et~al., 1999, \mn@doi [\apjs] {10.1086/313233}, \href
  {http://adsabs.harvard.edu/abs/1999ApJS..123....3L} {123, 3}

\bibitem[\protect\citeauthoryear{{Lelli}, {Verheijen}  \& {Fraternali}}{{Lelli}
  et~al.}{2014}]{Lelli2014b}
{Lelli} F.,  {Verheijen} M.,   {Fraternali} F.,  2014, \mn@doi [\aap]
  {10.1051/0004-6361/201322657}, \href
  {http://adsabs.harvard.edu/abs/2014A%26A...566A..71L} {566, A71}

\bibitem[\protect\citeauthoryear{{L{\'o}pez-S{\'a}nchez}, {Koribalski}, {van
  Eymeren}, {Esteban}, {Kirby}, {Jerjen}  \&
  {Lonsdale}}{{L{\'o}pez-S{\'a}nchez} et~al.}{2012}]{Lopez-Sanchez2012}
{L{\'o}pez-S{\'a}nchez} {\'A}.~R.,  {Koribalski} B.~S.,  {van Eymeren} J.,
  {Esteban} C.,  {Kirby} E.,  {Jerjen} H.,   {Lonsdale} N.,  2012, \mn@doi
  [\mnras] {10.1111/j.1365-2966.2011.19762.x}, \href
  {http://adsabs.harvard.edu/abs/2012MNRAS.419.1051L} {419, 1051}

\bibitem[\protect\citeauthoryear{{Ma}, {Hopkins}, {Faucher-Gigu{\`e}re},
  {Zolman}, {Muratov}, {Kere{\v s}}  \& {Quataert}}{{Ma} et~al.}{2016}]{Ma2016}
{Ma} X.,  {Hopkins} P.~F.,  {Faucher-Gigu{\`e}re} C.-A.,  {Zolman} N.,
  {Muratov} A.~L.,  {Kere{\v s}} D.,   {Quataert} E.,  2016, \mn@doi [\mnras]
  {10.1093/mnras/stv2659}, \href
  {http://adsabs.harvard.edu/abs/2016MNRAS.456.2140M} {456, 2140}

\bibitem[\protect\citeauthoryear{{Makarov}, {Prugniel}, {Terekhova}, {Courtois}
   \& {Vauglin}}{{Makarov} et~al.}{2014}]{Muratov2014}
{Makarov} D.,  {Prugniel} P.,  {Terekhova} N.,  {Courtois} H.,   {Vauglin} I.,
  2014, \mn@doi [\aap] {10.1051/0004-6361/201423496}, \href
  {http://adsabs.harvard.edu/abs/2014A%26A...570A..13M} {570, A13}

\bibitem[\protect\citeauthoryear{{Marigo}, {Girardi}, {Bressan}, {Groenewegen},
  {Silva}  \& {Granato}}{{Marigo} et~al.}{2008}]{Marigo2008}
{Marigo} P.,  {Girardi} L.,  {Bressan} A.,  {Groenewegen} M.~A.~T.,  {Silva}
  L.,   {Granato} G.~L.,  2008, \mn@doi [\aap] {10.1051/0004-6361:20078467},
  \href {http://adsabs.harvard.edu/abs/2008A%26A...482..883M} {482, 883}

\bibitem[\protect\citeauthoryear{{Marlowe}, {Heckman}, {Wyse}  \&
  {Schommer}}{{Marlowe} et~al.}{1995}]{Marlowe1995}
{Marlowe} A.~T.,  {Heckman} T.~M.,  {Wyse} R.~F.~G.,   {Schommer} R.,  1995,
  \mn@doi [\apj] {10.1086/175101}, \href
  {http://adsabs.harvard.edu/abs/1995ApJ...438..563M} {438, 563}

\bibitem[\protect\citeauthoryear{{Martin}}{{Martin}}{1998}]{Martin1998}
{Martin} C.~L.,  1998, \mn@doi [\apj] {10.1086/306219}, \href
  {http://adsabs.harvard.edu/abs/1998ApJ...506..222M} {506, 222}

\bibitem[\protect\citeauthoryear{{Martin}, {Kobulnicky}  \& {Heckman}}{{Martin}
  et~al.}{2002}]{Martin2002}
{Martin} C.~L.,  {Kobulnicky} H.~A.,   {Heckman} T.~M.,  2002, \mn@doi [\apj]
  {10.1086/341092}, \href {http://adsabs.harvard.edu/abs/2002ApJ...574..663M}
  {574, 663}

\bibitem[\protect\citeauthoryear{{Mart{\'{\i}}nez-Delgado}
  et~al.,}{{Mart{\'{\i}}nez-Delgado} et~al.}{2012}]{Martinez-Delgado2012}
{Mart{\'{\i}}nez-Delgado} D.,  et~al., 2012, \mn@doi [\apjl]
  {10.1088/2041-8205/748/2/L24}, \href
  {http://adsabs.harvard.edu/abs/2012ApJ...748L..24M} {748, L24}

\bibitem[\protect\citeauthoryear{{McQuinn} et~al.,}{{McQuinn}
  et~al.}{2010a}]{McQuinn2010a}
{McQuinn} K.~B.~W.,  et~al., 2010a, \mn@doi [\apj]
  {10.1088/0004-637X/721/1/297}, \href
  {http://adsabs.harvard.edu/abs/2010ApJ...721..297M} {721, 297}

\bibitem[\protect\citeauthoryear{{McQuinn} et~al.,}{{McQuinn}
  et~al.}{2010b}]{McQuinn2010b}
{McQuinn} K.~B.~W.,  et~al., 2010b, \mn@doi [\apj]
  {10.1088/0004-637X/724/1/49}, \href
  {http://adsabs.harvard.edu/abs/2010ApJ...724...49M} {724, 49}

\bibitem[\protect\citeauthoryear{{McQuinn}, {Skillman}, {Dalcanton}, {Cannon},
  {Dolphin}, {Holtzman}, {Weisz}  \& {Williams}}{{McQuinn}
  et~al.}{2012}]{McQuinn2012a}
{McQuinn} K.~B.~W.,  {Skillman} E.~D.,  {Dalcanton} J.~J.,  {Cannon} J.~M.,
  {Dolphin} A.~E.,  {Holtzman} J.,  {Weisz} D.~R.,   {Williams} B.~F.,  2012,
  \mn@doi [\apj] {10.1088/0004-637X/759/1/77}, \href
  {http://adsabs.harvard.edu/abs/2012ApJ...759...77M} {759, 77}

\bibitem[\protect\citeauthoryear{{McQuinn}, {Mitchell}  \&
  {Skillman}}{{McQuinn} et~al.}{2015a}]{McQuinn2015b}
{McQuinn} K.~B.~W.,  {Mitchell} N.~P.,   {Skillman} E.~D.,  2015a, \mn@doi
  [\apjs] {10.1088/0067-0049/218/2/29}, \href
  {http://adsabs.harvard.edu/abs/2015ApJS..218...29M} {218, 29}

\bibitem[\protect\citeauthoryear{{McQuinn} et~al.,}{{McQuinn}
  et~al.}{2015b}]{McQuinn2015f}
{McQuinn} K.~B.~W.,  et~al., 2015b, \mn@doi [\apjl]
  {10.1088/2041-8205/815/2/L17}, \href
  {http://adsabs.harvard.edu/abs/2015ApJ...815L..17M} {815, L17}

\bibitem[\protect\citeauthoryear{{Meiksin}}{{Meiksin}}{2016}]{Meiksin2016}
{Meiksin} A.,  2016, \mn@doi [\mnras] {10.1093/mnras/stw1460}, \href
  {http://adsabs.harvard.edu/abs/2016MNRAS.461.2762M} {461, 2762}

\bibitem[\protect\citeauthoryear{{Meurer} et~al.,}{{Meurer}
  et~al.}{2006}]{Meurer2006}
{Meurer} G.~R.,  et~al., 2006, \mn@doi [\apjs] {10.1086/504685}, \href
  {http://adsabs.harvard.edu/abs/2006ApJS..165..307M} {165, 307}

\bibitem[\protect\citeauthoryear{{Mewe}, {Gronenschild}  \& {van den
  Oord}}{{Mewe} et~al.}{1985}]{Mewe1985}
{Mewe} R.,  {Gronenschild} E.~H.~B.~M.,   {van den Oord} G.~H.~J.,  1985,
  \aaps, \href {http://adsabs.harvard.edu/abs/1985A%26AS...62..197M} {62, 197}

\bibitem[\protect\citeauthoryear{{Mewe}, {Lemen}  \& {van den Oord}}{{Mewe}
  et~al.}{1986}]{Mewe1986}
{Mewe} R.,  {Lemen} J.~R.,   {van den Oord} G.~H.~J.,  1986, \aaps, \href
  {http://adsabs.harvard.edu/abs/1986A%26AS...65..511M} {65, 511}

\bibitem[\protect\citeauthoryear{{Mineo}, {Gilfanov}  \& {Sunyaev}}{{Mineo}
  et~al.}{2012}]{Mineo2012}
{Mineo} S.,  {Gilfanov} M.,   {Sunyaev} R.,  2012, \mn@doi [\mnras]
  {10.1111/j.1365-2966.2012.21831.x}, \href
  {http://adsabs.harvard.edu/abs/2012MNRAS.426.1870M} {426, 1870}

\bibitem[\protect\citeauthoryear{{Mould} \& {Sakai}}{{Mould} \&
  {Sakai}}{2008}]{Mould2008}
{Mould} J.,  {Sakai} S.,  2008, \mn@doi [\apjl] {10.1086/592964}, \href
  {http://adsabs.harvard.edu/abs/2008ApJ...686L..75M} {686, L75}

\bibitem[\protect\citeauthoryear{{Muratov}, {Kere{\v s}},
  {Faucher-Gigu{\`e}re}, {Hopkins}, {Quataert}  \& {Murray}}{{Muratov}
  et~al.}{2015}]{Muratov2015}
{Muratov} A.~L.,  {Kere{\v s}} D.,  {Faucher-Gigu{\`e}re} C.-A.,  {Hopkins}
  P.~F.,  {Quataert} E.,   {Murray} N.,  2015, \mn@doi [\mnras]
  {10.1093/mnras/stv2126}, \href
  {http://adsabs.harvard.edu/abs/2015MNRAS.454.2691M} {454, 2691}

\bibitem[\protect\citeauthoryear{{Muratov} et~al.,}{{Muratov}
  et~al.}{2017}]{Muratov2017}
{Muratov} A.~L.,  et~al., 2017, \mn@doi [\mnras] {10.1093/mnras/stx667}, \href
  {http://adsabs.harvard.edu/abs/2017MNRAS.468.4170M} {468, 4170}

\bibitem[\protect\citeauthoryear{{Navarro}, {Eke}  \& {Frenk}}{{Navarro}
  et~al.}{1996}]{Navarro1996}
{Navarro} J.~F.,  {Eke} V.~R.,   {Frenk} C.~S.,  1996, \mn@doi [\mnras]
  {10.1093/mnras/283.3.72L}, \href
  {http://adsabs.harvard.edu/abs/1996MNRAS.283L..72N} {283, L72}

\bibitem[\protect\citeauthoryear{{Navarro}, {Frenk}  \& {White}}{{Navarro}
  et~al.}{1997}]{Navarro1997}
{Navarro} J.~F.,  {Frenk} C.~S.,   {White} S.~D.~M.,  1997, \mn@doi [\apj]
  {10.1086/304888}, \href {http://adsabs.harvard.edu/abs/1997ApJ...490..493N}
  {490, 493}

\bibitem[\protect\citeauthoryear{{O{\~n}orbe}, {Boylan-Kolchin}, {Bullock},
  {Hopkins}, {Kere{\v s}}, {Faucher-Gigu{\`e}re}, {Quataert}  \&
  {Murray}}{{O{\~n}orbe} et~al.}{2015}]{Onorbe2015}
{O{\~n}orbe} J.,  {Boylan-Kolchin} M.,  {Bullock} J.~S.,  {Hopkins} P.~F.,
  {Kere{\v s}} D.,  {Faucher-Gigu{\`e}re} C.-A.,  {Quataert} E.,   {Murray} N.,
   2015, \mn@doi [\mnras] {10.1093/mnras/stv2072}, \href
  {http://adsabs.harvard.edu/abs/2015MNRAS.454.2092O} {454, 2092}

\bibitem[\protect\citeauthoryear{{Oman} et~al.,}{{Oman}
  et~al.}{2015}]{Oman2015}
{Oman} K.~A.,  et~al., 2015, \mn@doi [\mnras] {10.1093/mnras/stv1504}, \href
  {http://adsabs.harvard.edu/abs/2015MNRAS.452.3650O} {452, 3650}

\bibitem[\protect\citeauthoryear{{Ott}, {Walter}, {Brinks}, {Van Dyk}, {Dirsch}
   \& {Klein}}{{Ott} et~al.}{2001}]{Ott2001}
{Ott} J.,  {Walter} F.,  {Brinks} E.,  {Van Dyk} S.~D.,  {Dirsch} B.,   {Klein}
  U.,  2001, \mn@doi [\aj] {10.1086/324101}, \href
  {http://adsabs.harvard.edu/abs/2001AJ....122.3070O} {122, 3070}

\bibitem[\protect\citeauthoryear{{Ott}, {Walter}  \& {Brinks}}{{Ott}
  et~al.}{2005a}]{Ott2005a}
{Ott} J.,  {Walter} F.,   {Brinks} E.,  2005a, \mn@doi [\mnras]
  {10.1111/j.1365-2966.2005.08862.x}, \href
  {http://adsabs.harvard.edu/abs/2005MNRAS.358.1423O} {358, 1423}

\bibitem[\protect\citeauthoryear{{Ott}, {Walter}  \& {Brinks}}{{Ott}
  et~al.}{2005b}]{Ott2005b}
{Ott} J.,  {Walter} F.,   {Brinks} E.,  2005b, \mn@doi [\mnras]
  {10.1111/j.1365-2966.2005.08863.x}, \href
  {http://adsabs.harvard.edu/abs/2005MNRAS.358.1453O} {358, 1453}

\bibitem[\protect\citeauthoryear{{Peeples}, {Werk}, {Tumlinson}, {Oppenheimer},
  {Prochaska}, {Katz}  \& {Weinberg}}{{Peeples} et~al.}{2014}]{Peeples2014}
{Peeples} M.~S.,  {Werk} J.~K.,  {Tumlinson} J.,  {Oppenheimer} B.~D.,
  {Prochaska} J.~X.,  {Katz} N.,   {Weinberg} D.~H.,  2014, \mn@doi [\apj]
  {10.1088/0004-637X/786/1/54}, \href
  {http://adsabs.harvard.edu/abs/2014ApJ...786...54P} {786, 54}

\bibitem[\protect\citeauthoryear{{Raymond} \& {Smith}}{{Raymond} \&
  {Smith}}{1977}]{Raymond1977}
{Raymond} J.~C.,  {Smith} B.~W.,  1977, \mn@doi [\apjs] {10.1086/190486}, \href
  {http://adsabs.harvard.edu/abs/1977ApJS...35..419R} {35, 419}

\bibitem[\protect\citeauthoryear{{Rich}, {Collins}, {Black}, {Longstaff},
  {Koch}, {Benson}  \& {Reitzel}}{{Rich} et~al.}{2012}]{Rich2012}
{Rich} R.~M.,  {Collins} M.~L.~M.,  {Black} C.~M.,  {Longstaff} F.~A.,  {Koch}
  A.,  {Benson} A.,   {Reitzel} D.~B.,  2012, \mn@doi [\nat]
  {10.1038/nature10837}, \href
  {http://adsabs.harvard.edu/abs/2012Natur.482..192R} {482, 192}

\bibitem[\protect\citeauthoryear{{Salpeter}}{{Salpeter}}{1955}]{Salpeter1955}
{Salpeter} E.~E.,  1955, \mn@doi [\apj] {10.1086/145971}, \href
  {http://adsabs.harvard.edu/abs/1955ApJ...121..161S} {121, 161}

\bibitem[\protect\citeauthoryear{{Sawala} et~al.,}{{Sawala}
  et~al.}{2016}]{Sawala2016}
{Sawala} T.,  et~al., 2016, \mn@doi [\mnras] {10.1093/mnras/stw145}, \href
  {http://adsabs.harvard.edu/abs/2016MNRAS.457.1931S} {457, 1931}

\bibitem[\protect\citeauthoryear{{Schlafly} \& {Finkbeiner}}{{Schlafly} \&
  {Finkbeiner}}{2011}]{Schlafly2011}
{Schlafly} E.~F.,  {Finkbeiner} D.~P.,  2011, \mn@doi [\apj]
  {10.1088/0004-637X/737/2/103}, \href
  {http://adsabs.harvard.edu/abs/2011ApJ...737..103S} {737, 103}

\bibitem[\protect\citeauthoryear{{Schlegel}, {Finkbeiner}  \&
  {Davis}}{{Schlegel} et~al.}{1998}]{Schlegel1998}
{Schlegel} D.~J.,  {Finkbeiner} D.~P.,   {Davis} M.,  1998, \mn@doi [\apj]
  {10.1086/305772}, \href {http://adsabs.harvard.edu/abs/1998ApJ...500..525S}
  {500, 525}

\bibitem[\protect\citeauthoryear{{Simon}, {Bolatto}, {Leroy}, {Blitz}  \&
  {Gates}}{{Simon} et~al.}{2005}]{Simon2005}
{Simon} J.~D.,  {Bolatto} A.~D.,  {Leroy} A.,  {Blitz} L.,   {Gates} E.~L.,
  2005, \mn@doi [\apj] {10.1086/427684}, \href
  {http://adsabs.harvard.edu/abs/2005ApJ...621..757S} {621, 757}

\bibitem[\protect\citeauthoryear{{Simpson}, {Hunter}  \& {Knezek}}{{Simpson}
  et~al.}{2005}]{Simpson2005}
{Simpson} C.~E.,  {Hunter} D.~A.,   {Knezek} P.~M.,  2005, \mn@doi [\aj]
  {10.1086/426364}, \href {http://adsabs.harvard.edu/abs/2005AJ....129..160S}
  {129, 160}

\bibitem[\protect\citeauthoryear{{Smith}, {Brickhouse}, {Liedahl}  \&
  {Raymond}}{{Smith} et~al.}{2001}]{Smith2001}
{Smith} R.~K.,  {Brickhouse} N.~S.,  {Liedahl} D.~A.,   {Raymond} J.~C.,  2001,
  \mn@doi [\apjl] {10.1086/322992}, \href
  {http://adsabs.harvard.edu/abs/2001ApJ...556L..91S} {556, L91}

\bibitem[\protect\citeauthoryear{{Spitoni}, {Calura}, {Matteucci}  \&
  {Recchi}}{{Spitoni} et~al.}{2010}]{Spitoni2010}
{Spitoni} E.,  {Calura} F.,  {Matteucci} F.,   {Recchi} S.,  2010, \mn@doi
  [\aap] {10.1051/0004-6361/200913799}, \href
  {http://adsabs.harvard.edu/abs/2010A%26A...514A..73S} {514, A73}

\bibitem[\protect\citeauthoryear{{Stil} \& {Israel}}{{Stil} \&
  {Israel}}{2002}]{Stil2002}
{Stil} J.~M.,  {Israel} F.~P.,  2002, \mn@doi [\aap]
  {10.1051/0004-6361:20020953}, \href
  {http://adsabs.harvard.edu/abs/2002A%26A...392..473S} {392, 473}

\bibitem[\protect\citeauthoryear{{Strickland} \& {Stevens}}{{Strickland} \&
  {Stevens}}{2000}]{Strickland2000a}
{Strickland} D.~K.,  {Stevens} I.~R.,  2000, \mn@doi [\mnras]
  {10.1046/j.1365-8711.2000.03391.x}, \href
  {http://adsabs.harvard.edu/abs/2000MNRAS.314..511S} {314, 511}

\bibitem[\protect\citeauthoryear{{Suchkov}, {Balsara}, {Heckman}  \&
  {Leitherer}}{{Suchkov} et~al.}{1994}]{Suchkov1994}
{Suchkov} A.~A.,  {Balsara} D.~S.,  {Heckman} T.~M.,   {Leitherer} C.,  1994,
  \mn@doi [\apj] {10.1086/174427}, \href
  {http://adsabs.harvard.edu/abs/1994ApJ...430..511S} {430, 511}

\bibitem[\protect\citeauthoryear{{Summers}, {Stevens}, {Strickland}  \&
  {Heckman}}{{Summers} et~al.}{2003}]{Summers2003}
{Summers} L.~K.,  {Stevens} I.~R.,  {Strickland} D.~K.,   {Heckman} T.~M.,
  2003, \mn@doi [\mnras] {10.1046/j.1365-8711.2003.06590.x}, \href
  {http://adsabs.harvard.edu/abs/2003MNRAS.342..690S} {342, 690}

\bibitem[\protect\citeauthoryear{{Summers}, {Stevens}, {Strickland}  \&
  {Heckman}}{{Summers} et~al.}{2004}]{Summers2004}
{Summers} L.~K.,  {Stevens} I.~R.,  {Strickland} D.~K.,   {Heckman} T.~M.,
  2004, \mn@doi [\mnras] {10.1111/j.1365-2966.2004.07749.x}, \href
  {http://adsabs.harvard.edu/abs/2004MNRAS.351....1S} {351, 1}

\bibitem[\protect\citeauthoryear{{Tremonti} et~al.,}{{Tremonti}
  et~al.}{2004}]{Tremonti2004}
{Tremonti} C.~A.,  et~al., 2004, \mn@doi [\apj] {10.1086/423264}, \href
  {http://adsabs.harvard.edu/abs/2004ApJ...613..898T} {613, 898}

\bibitem[\protect\citeauthoryear{{T{\"u}llmann} et~al.,}{{T{\"u}llmann}
  et~al.}{2011}]{Tullmann2011}
{T{\"u}llmann} R.,  et~al., 2011, \mn@doi [\apjs] {10.1088/0067-0049/193/2/31},
  \href {http://adsabs.harvard.edu/abs/2011ApJS..193...31T} {193, 31}

\bibitem[\protect\citeauthoryear{{Walker} \& {Pe{\~n}arrubia}}{{Walker} \&
  {Pe{\~n}arrubia}}{2011}]{Walker2011}
{Walker} M.~G.,  {Pe{\~n}arrubia} J.,  2011, \mn@doi [\apj]
  {10.1088/0004-637X/742/1/20}, \href
  {http://adsabs.harvard.edu/abs/2011ApJ...742...20W} {742, 20}

\bibitem[\protect\citeauthoryear{{Warren} et~al.,}{{Warren}
  et~al.}{2011}]{Warren2011}
{Warren} S.~R.,  et~al., 2011, \mn@doi [\apj] {10.1088/0004-637X/738/1/10},
  \href {http://adsabs.harvard.edu/abs/2011ApJ...738...10W} {738, 10}

\bibitem[\protect\citeauthoryear{{Westmoquette}, {Smith}  \&
  {Gallagher}}{{Westmoquette} et~al.}{2008}]{Westmoquette2008}
{Westmoquette} M.~S.,  {Smith} L.~J.,   {Gallagher} J.~S.,  2008, \mn@doi
  [\mnras] {10.1111/j.1365-2966.2007.12628.x}, \href
  {http://adsabs.harvard.edu/abs/2008MNRAS.383..864W} {383, 864}

\bibitem[\protect\citeauthoryear{{Zahid}, {Dima}, {Kewley}, {Erb}  \&
  {Dav{\'e}}}{{Zahid} et~al.}{2012}]{Zahid2012}
{Zahid} H.~J.,  {Dima} G.~I.,  {Kewley} L.~J.,  {Erb} D.~K.,   {Dav{\'e}} R.,
  2012, \mn@doi [\apj] {10.1088/0004-637X/757/1/54}, \href
  {http://adsabs.harvard.edu/abs/2012ApJ...757...54Z} {757, 54}

\bibitem[\protect\citeauthoryear{{de Blok}}{{de Blok}}{2010}]{deBlok2010}
{de Blok} W.~J.~G.,  2010, \mn@doi [Advances in Astronomy]
  {10.1155/2010/789293}, \href
  {http://adsabs.harvard.edu/abs/2010AdAst2010E...5D} {2010, 789293}

\bibitem[\protect\citeauthoryear{{van de Voort}, {Quataert}, {Hopkins},
  {Faucher-Gigu{\`e}re}, {Feldmann}, {Kere{\v s}}, {Chan}  \& {Hafen}}{{van de
  Voort} et~al.}{2016}]{vandeVoort2016}
{van de Voort} F.,  {Quataert} E.,  {Hopkins} P.~F.,  {Faucher-Gigu{\`e}re}
  C.-A.,  {Feldmann} R.,  {Kere{\v s}} D.,  {Chan} T.~K.,   {Hafen} Z.,  2016,
  \mn@doi [\mnras] {10.1093/mnras/stw2322}, \href
  {http://adsabs.harvard.edu/abs/2016MNRAS.463.4533V} {463, 4533}

\makeatother
\end{thebibliography}

\bsp	
\label{lastpage}
\end{document}